\documentclass[zpreprint,zbstnp]{zeus_paper}
\usepackage[english]{babel}

\newcommand{\ZcoosysB}{%
The ZEUS coordinate system is a right-handed Cartesian system, with the $Z$
axis pointing in the proton beam direction, referred to as the ``forward
direction'', and the $X$ axis pointing left towards the centre of HERA.
The coordinate origin is at the nominal interaction point.\xspace}
\newcommand{\Zpsrap}{%
The pseudorapidity is defined as $\eta=-\ln\left(\tan\frac{\theta}{2}\right)$,
where the polar angle, $\theta$, is measured with respect to the proton beam
direction.\xspace}

\newcommand{\Zdetdesc}{%
A detailed description of the ZEUS detector can be found 
elsewhere~\cite{zeus:1993:bluebook}. A brief outline of the 
components that are most relevant for this analysis is given
below.\xspace}
\newcommand{\Zctddesc}[1]{%
Charged particles are tracked in the central tracking detector (CTD)~\citeCTD,
which operates in a magnetic field of $1.43\Tesla$ provided by a thin 
superconducting coil. The CTD consists of 72~cylindrical drift chamber 
layers, organized in 9~superlayers covering the polar-angle#1 region 
\mbox{$15^\circ<\theta<164^\circ$}. The transverse-momentum resolution for
full-length tracks is $\sigma(p_T)/p_T=0.0058p_T\oplus0.0065\oplus0.0014/p_T$,
with $p_T$ in $\Gev$.}
\newcommand{\Zcaldesc}{%
The high-resolution uranium--scintillator calorimeter (CAL)~\citeCAL consists 
of three parts: the forward (FCAL), the barrel (BCAL) and the rear (RCAL)
calorimeters. Each part is subdivided transversely into towers and
longitudinally into one electromagnetic section (EMC) and either one (in RCAL)
or two (in BCAL and FCAL) hadronic sections (HAC). The smallest subdivision of
the calorimeter is called a cell.  The CAL energy resolutions, as measured under
test-beam conditions, are $\sigma(E)/E=0.18/\sqrt{E}$ for electrons and
$\sigma(E)/E=0.35/\sqrt{E}$ for hadrons ($E$ in $\Gev$).}
\chardef\usc=95
\chardef\til=126
\catcode`\@=11 % @ signs are now treated as letters
\DeclareRobustCommand\xdotspace{\futurelet\@let@token\@xdotspace}
\def\@xdotspace{%
  \ifx\@let@token.\else
  \ifx\@let@token\bgroup.\else
  \ifx\@let@token\egroup.\else
  \ifx\@let@token\/.\else
  \ifx\@let@token\ .\else
  \ifx\@let@token~.\else
  \ifx\@let@token!.\else
  \ifx\@let@token,.\else
  \ifx\@let@token:.\else
  \ifx\@let@token;.\else
  \ifx\@let@token?.\else
  \ifx\@let@token/.\else
  \ifx\@let@token'.\else
  \ifx\@let@token).\else
  \ifx\@let@token-.\else
  \ifx\@let@token\@xobeysp.\else
  \ifx\@let@token\space.\else
  \ifx\@let@token\@sptoken.\else
   .\space
   \fi\fi\fi\fi\fi\fi\fi\fi\fi\fi\fi\fi\fi\fi\fi\fi\fi\fi}
\catcode`\@=12 % @ signs are no longer letters

\newcommand{\stru}[2]{%
   \relax\ifmmode\hbox{\vrule height#1 depth#2 width0pt}%
   \else\vrule height#1 depth#2 width0pt\fi}
\newcommand{\Ronum}[1]{\uppercase\expandafter{\romannumeral#1}}
\newcommand{\ronum}[1]{\expandafter{\romannumeral#1}}
\DeclareRobustCommand{\LaTeXZ}{%
  \LaTeX\kern-.05em4\kern-.1em
  {\raisebox{-0.2ex}{$\scriptstyle\text{ZEUS}$}}\xspace}
\DeclareMathAlphabet{\mathbf}{OT1}{cmr}{bx}{sl}
\newcommand{\eVdist}{\kern-0.06667em}

\newcommand{\Gev}{{\text{Ge}\eVdist\text{V\/}}}

\newcommand{\gev}{{\,\text{Ge}\eVdist\text{V\/}}}

\newcommand{\pb}{\,\text{pb}}

\newcommand{\pbi}{\,\text{pb}^{-1}}

\newcommand{\met}{\,\text{m}}

\newcommand{\cm}{\,\text{cm}}

\newcommand{\ns}{\,\text{ns}}

\newcommand{\Tesla}{\,\text{T}}

\newcommand{\slashfrac}[2]{%
  \raisebox{0.5ex}{\ensuremath #1}\kern-0.12em/\kern-0.08em
  \raisebox{-.8ex}{\ensuremath #2}}
\newcommand{\sqr}[3]{%
    {\vcenter{\hrule height.#3ex\hbox{\vrule width.#2ex height#1ex
     \kern#1ex\vrule width.#3ex}\hrule height.#2ex}}}

\newcommand{\widebar}[1]{%
   \mkern1.5mu\overline{\mkern-1.5mu#1\mkern-1.mu}\mkern1.mu}
\catcode`\@=11 % @ signs are now treated as letters
\newcommand{\parenbar}{\mathpalette\p@renb@r}
\def\p@renb@r#1#2{\vbox{%
  \ifx#1\scriptscriptstyle \dimen@.7em\dimen@ii.2em\else
  \ifx#1\scriptstyle \dimen@.8em\dimen@ii.25em\else
  \dimen@1em\dimen@ii.4em\fi\fi \offinterlineskip
  \ialign{\hfill##\hfill\cr
    \vbox{\hrule width\dimen@ii}\cr
    \noalign{\vskip-.3ex}%
    \hbox to\dimen@{$\mathchar300\hfil\mathchar301$}\cr
    \noalign{\vskip-.3ex}%
    $#1#2$\cr}}}
\catcode`\@=12 % @ signs are no longer letters

\newcommand{\pbar}{\widebar{p}}

\newcommand{\cbar}{\widebar{c}}
\newcommand{\bbar}{\widebar{b}}

\newcommand{\rnge}{\hbox{$\,\text{--}\,$}}

\newcommand{\IP}{{\rm I$\kern-0.01667em$P}\xspace}

\mathchardef\qsm=63
\mathchardef\pls=43
\mathchardef\mns=512
\mathchardef\plm=518
\mathchardef\eql=61
\mathchardef\smallleft=300
\mathchardef\smallright=301
\mathchardef\les=316
\mathchardef\gre=318
\mathchardef\leq=532
\mathchardef\grq=533
\catcode`\@=11 % @ signs are now treated as letters
\newcounter{pict@width}
\newcounter{pict@height}
\newlength{\pict@scale}
\newcommand{\psfigadd}[4]{%
\setcounter{pict@width}{1*\ratio{#2+\pict@scale/2}{\pict@scale}}
\setcounter{pict@height}{1*\ratio{#3+\pict@scale/2}{\pict@scale}}
\setlength{\unitlength}{\pict@scale}
\hbox to #2{\hspace{-\fill}\begin{picture}(\thepict@width,\thepict@height)
\put(0,0){\psfig{figure=#1,width=#2,height=#3,clip=}}
\SetScale{0.283466457}
\SetWidth{1.763889}
{#4}
\end{picture}}
}
\newcounter{pict@widthfst}
\newcounter{pict@widthscd}
\newcounter{pict@widthtot}
\newcommand{\psfigaddtwo}[7]{%
\setcounter{pict@widthfst}{1*\ratio{#2+\pict@scale/2}{\pict@scale}}
\setcounter{pict@widthscd}{1*\ratio{#2+#4+\pict@scale/2}{\pict@scale}}
\setcounter{pict@widthtot}{1*\ratio{#2+#4+#6+\pict@scale/2}{\pict@scale}}
\setcounter{pict@height}{1*\ratio{#3+\pict@scale/2}{\pict@scale}}
\setlength{\unitlength}{\pict@scale}
\hbox{\hspace{-\fill}\begin{picture}(\thepict@widthtot,\thepict@height)
\put(0,0){\psfig{figure=#1,width=#2,height=#3,clip=}}
\put(\thepict@widthscd,0){\psfig{figure=#5,width=#6,height=#3,clip=}}
\SetScale{0.283466457}
\SetWidth{1.763889}
{#7}
\end{picture}}
}
\newcommand{\psfigror}[4]{%
\setcounter{pict@width}{1*\ratio{#2+\pict@scale/2}{\pict@scale}}
\setcounter{pict@height}{1*\ratio{#3+\pict@scale/2}{\pict@scale}}
\setlength{\unitlength}{\pict@scale}
\hbox{\begin{picture}(\thepict@width,\thepict@height)
\put(0,\thepict@height){\psfig{figure=#1,width=#3,height=#2,clip=,angle=270}}
\SetScale{0.283466457}
\SetWidth{1.763889}
{#4}
\end{picture}}
}
\newcommand{\psfigrol}[4]{%
\setcounter{pict@width}{1*\ratio{#2+\pict@scale/2}{\pict@scale}}
\setcounter{pict@height}{1*\ratio{#3+\pict@scale/2}{\pict@scale}}
\setlength{\unitlength}{\pict@scale}
\hbox{\begin{picture}(\thepict@width,\thepict@height)
\put(0,0){\psfig{figure=#1,width=#3,height=#2,clip=,angle=90}}
\SetScale{0.283466457}
\SetWidth{1.763889}
{#4}
\end{picture}}
}
\catcode`\@=12 % @ signs are no longer letters
\newlength\listtextwidth

\catcode`\@=11 % @ signs are now treated as letters
\newlength{\@tabfninsert}
\newlength{\@tabfnwidth}
\newcommand{\tabfootnote}[2]{%
  \setlength{\@tabfninsert}{0.8em}
  \setlength{\@tabfnwidth}{\textwidth}
  \addtolength{\@tabfnwidth}{-\@tabfninsert}
  \addtolength{\@tabfnwidth}{-0.4em}
  \noindent\makebox[\@tabfninsert][r]{\footnotesize$^{#1}$\hfil}\hfill%
  \parbox[t]{\@tabfnwidth}{\footnotesize #2\hfill}}
\catcode`\@=12 % @ signs are no longer letters

\def\citeCTD{{\cite{%
nim:a279:290,*npps:b32:181,*nim:a338:254%
}}\xspace}
\def\citeCAL{{\cite{%
nim:a309:77,*nim:a309:101,*nim:a321:356,*nim:a336:23%
}}\xspace}
\renewcommand{\Zctddesc}[1]{%
Charged particles were tracked in the central tracking detector (CTD)~\citeCTD,
which operated in a magnetic field of $1.43\Tesla$ provided by a thin 
superconducting coil. The CTD consisted of 72~cylindrical drift chamber 
layers, organised in nine superlayers covering the polar-angle#1 region 
\mbox{$15^\circ<\theta<164^\circ$}. The transverse-momentum resolution for
full-length tracks is $\sigma(p_T)/p_T=0.0058p_T\oplus0.0065\oplus0.0014/p_T$,
with $p_T$ in $\Gev$.}
\renewcommand{\Zcaldesc}{%
The high-resolution uranium--scintillator calorimeter (CAL)~\citeCAL consisted
of three parts: the forward (FCAL), the barrel (BCAL) and the rear (RCAL)
calorimeters. Each part was subdivided transversely into towers and
longitudinally into one electromagnetic section and either one (in RCAL)
or two (in BCAL and FCAL) hadronic sections. The smallest subdivision of
the calorimeter is called a cell.  The CAL energy resolutions, as measured under
test-beam conditions, are $\sigma(E)/E=0.18/\sqrt{E}$ for electrons and
$\sigma(E)/E=0.35/\sqrt{E}$ for hadrons, with $E$ in $\Gev$.}
\newcommand{\ZcoosysfnBetaphi}{\footnote{\ZcoosysB\Zpsrap{} The
    azimuthal angle, $\phi$, is measured with respect to the $X$ axis.}}

\newcommand{\Qsq}{\ensuremath{{Q^{2}}}}
\newcommand{\pT}{\ensuremath{p_{T}}}
\newcommand{\ET}{\ensuremath{E_{T}}}
\newcommand{\Ejet}{\ensuremath{E^{\textrm{jet}}}}
\newcommand{\ETjet}{\ensuremath{E_{T}^{\textrm{jet}}}}
\newcommand{\ETjeti}{\ensuremath{E_{T}^{\textrm{jet 1}}}}

\newcommand{\ETejet}{\ensuremath{E_{T}^{\textrm{e jet}}}}
\newcommand{\pjet}{\ensuremath{p^{\textrm{jet}}}}
\newcommand{\qjet}{\ensuremath{q^{\textrm{jet}}}}
\newcommand{\pTjet}{\ensuremath{p_{T}^{\textrm{jet}}}}
\newcommand{\etajet}{\ensuremath{\eta^{\textrm{jet}}}}
\newcommand{\etajeti}{\ensuremath{\eta^{\textrm{jet 1}}}}

\newcommand{\etaejet}{\ensuremath{\eta_{T}^{\textrm{e jet}}}}
\newcommand{\phijet}{\ensuremath{\phi^{\textrm{jet}}}}
\newcommand{\pTrel}{\ensuremath{p_{\perp}^{\textrm{rel}}}}
\newcommand{\pTb}{\ensuremath{p_{T}^{b}}}
\newcommand{\pTe}{\ensuremath{p_{T}^{e}}}
\newcommand{\etae}{\ensuremath{\eta^{e}}}
\newcommand{\pTmiss}{\ensuremath{\vec{\not\!p}_{T}}}
\newcommand{\Dphi}{\ensuremath{\Delta\phi}}

\newcommand{\xgamobs}{\ensuremath{x_{\gamma}^{\textrm{obs}}}}
\newcommand{\Chad}{\ensuremath{C^{\mathrm{had}}}}
\newcommand{\Chadb}{\ensuremath{C^{\mathrm{had}}_{b}}}
\newcommand{\Chadc}{\ensuremath{C^{\mathrm{had}}_{c}}}
\newcommand{\yJB}{\ensuremath{y_{\mathrm{JB}}}}
\newcommand{\eSL}{\ensuremath{e_{\mathrm{SL}}}}
\newcommand{\bbbar}{\ensuremath{b\bbar}}
\newcommand{\ccbar}{\ensuremath{c\cbar}}
\newcommand{\bpbbar}{\ensuremath{b(\bbar)}}
\newcommand{\cpcbar}{\ensuremath{c(\cbar)}}
\newcommand{\Jpsi}{\ensuremath{J/\psi}}

\newcommand{\signlob}{\ensuremath{\sigma_{b}^{\textrm{NLO}}}}
\newcommand{\signloc}{\ensuremath{\sigma_{c}^{\textrm{NLO}}}}
\newcommand{\sigvisb}{\ensuremath{\sigma_{b}^{\textrm{vis}}}}
\newcommand{\sigvisc}{\ensuremath{\sigma_{c}^{\textrm{vis}}}}

\newcommand{\dEdx}{\ensuremath{\mathrm{d}E/\mathrm{d}x}}
\newcommand{\EEMC}{\ensuremath{E_{\mathrm{EMC}}}}
\newcommand{\ECAL}{\ensuremath{E_{\mathrm{CAL}}}}
\newcommand{\EMCfrac}{\ensuremath{E_{\mathrm{EMC}}/E_{\mathrm{CAL}}}}
\newcommand{\ntrunc}{\ensuremath{n_{\mathrm{trunc}}}}
\newcommand{\Ncand}{\ensuremath{N_{\mathrm{cand}}}}
\newcommand{\stat}{\ensuremath{\textrm{stat.}}}
\newcommand{\syst}{\ensuremath{\textrm{syst.}}}
\newcommand{\PYTHIA}{\textsc{Pythia}}
\begin{document}
\prepnum{DESY--08--056}

\title{Beauty photoproduction using decays into electrons at HERA}
                    
\author{ZEUS Collaboration}
\date{29 May 2008}

\abstract{Photoproduction of beauty quarks in events with two jets and
  an electron associated with one of the jets has been studied with
  the ZEUS detector at HERA using an integrated luminosity of
  $120\pbi$.  The fractions of events containing $b$ quarks, and also
  of events containing $c$ quarks, were extracted from a likelihood
  fit using variables sensitive to electron identification as well as
  to semileptonic decays.  Total and differential cross sections for
  beauty and charm production were measured and compared with
  next-to-leading-order QCD calculations and Monte Carlo models.}

\makezeustitle
\def\3{\ss}                                                                                        
\pagenumbering{Roman}

\begin{center}                                                                                     
{                      \Large  The ZEUS Collaboration              }                               
\end{center}                                                                                       
  S.~Chekanov,                                                                                     
  M.~Derrick,                                                                                      
  S.~Magill,                                                                                       
  B.~Musgrave,                                                                                     
  D.~Nicholass$^{   1}$,                                                                           
  \mbox{J.~Repond},                                                                                
  R.~Yoshida\\                                                                                     
 {\it Argonne National Laboratory, Argonne, Illinois 60439-4815, USA}~$^{n}$                       
\par \filbreak                                                                                     
  M.C.K.~Mattingly \\                                                                              
 {\it Andrews University, Berrien Springs, Michigan 49104-0380, USA}                               
\par \filbreak                                                                                     
  P.~Antonioli,                                                                                    
  G.~Bari,                                                                                         
  L.~Bellagamba,                                                                                   
  D.~Boscherini,                                                                                   
  A.~Bruni,                                                                                        
  G.~Bruni,                                                                                        
  F.~Cindolo,                                                                                      
  M.~Corradi,                                                                                      
\mbox{G.~Iacobucci},                                                                               
  A.~Margotti,                                                                                     
  R.~Nania,                                                                                        
  A.~Polini\\                                                                                      
  {\it INFN Bologna, Bologna, Italy}~$^{e}$
% ...
\newpage                                                                                           
\enlargethispage{5cm}                                                                              
$^{\    1}$ also affiliated with University College London, UK \\                                  
% ...
$^{\    5}$ This work was supported in part by the Marie Curie Actions Transfer of Knowledge       
project COCOS (contract MTKD-CT-2004-517186)\\                                                     
% ...
%                                                                                                  
% \par         % if index listing & table fit to 1 page, put gap here                              
\newpage   % alternatively: go to newpage, if page is too small                                    
                                                           %                                       
% \institute_references_start    % do not touch or move this line !                                
                                                           %                                       
\begin{tabular}[h]{rp{14cm}}                                                                       
$^{a}$ &  supported by the Natural Sciences and Engineering Research Council of Canada (NSERC) \\  
$^{b}$ &  supported by the German Federal Ministry for Education and Research (BMBF), under        
          contract numbers 05 HZ6PDA, 05 HZ6GUA, 05 HZ6VFA and 05 HZ4KHA\\                         
$^{c}$ &  supported in part by the MINERVA Gesellschaft f\"ur Forschung GmbH, the Israel Science   
          Foundation (grant no. 293/02-11.2) and the U.S.-Israel Binational Science Foundation \\  
$^{d}$ &  supported by the Israel Science Foundation\\                                             
$^{e}$ &  supported by the Italian National Institute for Nuclear Physics (INFN) \\                
$^{f}$ &  supported by the Japanese Ministry of Education, Culture, Sports, Science and Technology 
          (MEXT) and its grants for Scientific Research\\                                          
$^{g}$ &  supported by the Korean Ministry of Education and Korea Science and Engineering          
          Foundation\\                                                                             
$^{h}$ &  supported by the Netherlands Foundation for Research on Matter (FOM)\\                   
$^{i}$ &  supported by the Polish State Committee for Scientific Research, project no.             
          DESY/256/2006 - 154/DES/2006/03\\                                                        
$^{j}$ &  partially supported by the German Federal Ministry for Education and Research (BMBF)\\   
$^{k}$ &  supported by RF Presidential grant N 8122.2006.2 for the leading                         
          scientific schools and by the Russian Ministry of Education and Science through its      
          grant for Scientific Research on High Energy Physics\\                                   
$^{l}$ &  supported by the Spanish Ministry of Education and Science through funds provided by     
          CICYT\\                                                                                  
$^{m}$ &  supported by the Science and Technology Facilities Council, UK\\                         
$^{n}$ &  supported by the US Department of Energy\\                                               
$^{o}$ &  supported by the US National Science Foundation. Any opinion,                            
findings and conclusions or recommendations expressed in this material                             
are those of the authors and do not necessarily reflect the views of the                           
National Science Foundation.\\                                                                     
$^{p}$ &  supported by the Polish Ministry of Science and Higher Education                         
as a scientific project (2006-2008)\\                                                              
$^{q}$ &  supported by FNRS and its associated funds (IISN and FRIA) and by an Inter-University    
          Attraction Poles Programme subsidised by the Belgian Federal Science Policy Office\\     
$^{r}$ &  supported by the Malaysian Ministry of Science, Technology and                           
Innovation/Akademi Sains Malaysia grant SAGA 66-02-03-0048\\                                       
\end{tabular}                                                                                      
                                                           %                                       
% \institute_references_end     % do not touch or move this line !                                 
\clearpage

%------------------------------------------------------------------------------
%       Text
%------------------------------------------------------------------------------
\pagenumbering{arabic} 
\pagestyle{plain}
% ----------------------------------------------------------------------------
%       Introduction
% ----------------------------------------------------------------------------
\section{Introduction}
\label{sec:int}

The production of heavy quarks in $ep$ collisions at HERA is an
important testing ground for perturbative Quantum Chromodynamics
(pQCD) since the large $b$-quark and $c$-quark masses provide a hard
scale that allows perturbative calculations.  When \Qsq{}, the
negative squared four-momentum exchanged at the electron or
positron\footnote{Hereafter unless explicitly stated both electrons
  and positrons are referred to as electrons.} vertex, is small, the
reactions 
$ep \rightarrow e \, b\bbar \, X$ and 
$ep \rightarrow e \, c\cbar \, X$ 
can be considered as a photoproduction process in which a
quasi-real photon, emitted by the incoming electron interacts with the
proton.

The corresponding leading-order (LO) QCD processes
are the direct-photon process, in which the quasi-real photon enters directly
in the hard interaction, 
and the resolved-photon process, in which the photon acts as a source of partons which
take part in the hard interaction.
For heavy-quark transverse momenta comparable to the quark mass,
next-to-leading-order (NLO) QCD calculations in which the massive
quark is generated
dynamically~\cite{pr:d45:3986,np:b454:3,*pl:b348:633}
are expected to provide reliable
predictions for the photoproduction cross sections.
 
Beauty and charm quark production cross sections have been measured using
several different methods by both the
ZEUS\cite{epj:c50:1434,pl:b599:173,pr:d70:012008,epj:c18:625,%
pl:b649:111,desy-07-052,np:b729:492,epj:c44:351,pr:d69:012004,pl:b565:87,epj:c12:35,pl:b481:213,epj:c6:67,pl:b401:192,pl:b407:402,pl:b349:225}
and the H1\cite{epj:c47:597,epj:c45:23,pl:b621:56,epj:c40:349,epj:c41:453,pl:b467:156,%
epj:c51:271,epj:c50:251,epj:c38:447,pl:b528:199,np:b545:21,np:b472:32} collaborations. 
Both the deep
inelastic scattering (DIS) and photoproduction measurements
are reasonably well described by NLO QCD predictions.

Most of the previous measurements of $b$-quark production used muons to tag
semileptonic decays of the $B$ hadrons.  The identification of
electrons close to jets is more difficult than for muons, but the electrons can be
identified down to lower momenta. 
A first measurement of $b$-quark photoproduction from semileptonic
decays to electrons ($e^{-}$) was presented in a previous
publication\cite{epj:c18:625}, which used the $e^{+}p$ collision data
from the 1996--1997 running period corresponding to an integrated
luminosity of $38\pbi$.  This paper presents an extension of this
measurement exploiting semileptonic decays to electrons as well as to
positrons for data taken with both $e^{-}p$ and $e^{+}p$ collisions
using three times the integrated luminosity.
The production of electrons from semileptonic decays (\eSL), in events with
at least two jets ($jj$) in photoproduction, $ep \rightarrow e \, b \bbar \,
X \rightarrow e \, jj \, \eSL \, X'$, was measured in the kinematic
range $\Qsq < 1\gev^2$ and $140\gev < W_{\gamma p} < 280\gev$, where
$W_{\gamma p}$ is the centre-of-mass energy of the $\gamma p$ system.
The likelihood method used to extract the $b$-quark cross sections
also allowed the corresponding $c$-quark cross sections to be
extracted. This paper provides a complementary study to the measurements using muon decays.

% ----------------------------------------------------------------------------
%       Experimental set-up
% ----------------------------------------------------------------------------
\section{Experimental set-up}
\label{sec:exp}

This analysis was performed with data taken from 1996 to 2000, when
HERA collided electrons or positrons with energy $E_e=27.5\gev$ with
protons of energy $E_p=820\gev$ (1996--1997) or 920\gev (1998--2000).
The corresponding integrated luminosities are $38.6 \pm 0.6\pbi$ at
centre-of-mass energy $\sqrt{s} = 300\gev$, and $81.6 \pm 1.8\pbi$ at
$\sqrt{s} = 318\gev$.

\Zdetdesc 

\Zctddesc\ZcoosysfnBetaphi\   
The pulse height of the sense wires was read out in order to estimate
the ionisation energy loss per unit length, \dEdx{} (see Section~\ref{sec:dedx}).

\Zcaldesc

The luminosity was measured from the rate of the bremsstrahlung process $ep
\rightarrow e\gamma p$, where the photon was measured in a lead--scintillator
calorimeter\cite{desy-92-066,*zfp:c63:391,*acpp:b32:2025} placed in the HERA
tunnel at $Z = -107\met$.

%------------------------------------------------------------------------------
% dE/dx subsection
%------------------------------------------------------------------------------
\section{\boldmath \dEdx{} Measurement}
\label{sec:dedx}

A central tool for this analysis was the \dEdx{} measurement from
the CTD. The pulse height of the signals on the sense wires was
used to measure the specific ionisation.  This pulse
height was corrected for a number of effects\cite{thesis:bartsch:2007}.
Such as a factor $1/\sin \theta$ due to the projection of the track onto
the direction of the signal wire,
the space-charge effect caused by the overlap of the ionisation clouds
in the avalanche, and the dependence of the pulse shape on the track topology.
An additional correction was needed for hits
close to the end-plates of the CTD.  
If a hit followed a previous one on the same wire within 100\ns,
its pulse could be distorted: such hits were
rejected.
The event topology was used to identify additional double hits that
could not be resolved; the \dEdx{} measurement was corrected accordingly.

The \dEdx{} value of a track was calculated as the truncated mean
value of the individual measurements, corrected as discussed
above, after rejecting the lowest 10\%
and the highest 30\% of the measurements.
Hits where the measured pulse height was in saturation were always rejected in forming the mean.
Corrections were applied
for the finite number of hits
and whenever more than 30\% of
the hits were saturated.
The corrected \dEdx{} measurement was normalised in units of mip
(minimum ionising particles) such that the minimum of the \dEdx{}
distribution was 1.0 mip. Electrons are expected to have a mean value of
about 1.4 mip in the momentum range studied here.

Different samples of identified particles were used to calibrate and
validate the \dEdx{} measurement.
The samples used for calibration were:
\begin{itemize}
\item $e^{\pm}$ from photon conversions, $J/\psi$ decays and DIS electrons;
\item $\pi^{\pm}$ from $K^0$ decays with $0.4\gev < p <1\gev$, where
  $p$ is the measured track momentum.
\end{itemize}
The samples used for validation were: 
\begin{itemize}
\item $\pi^{\pm}$ from $K^0$ outside the momentum range used for
  the calibration sample,\\
  as well as $\pi^{\pm}$ from $\rho^0$, $\Lambda$ and $D^*$ decays;
\item $K^{\pm}$ from
  $\phi^0$ and $D^*$ decays;
\item $p,\pbar$ from $\Lambda$ decays;
\item cosmic $\mu^{\pm}$. 
\end{itemize}
Typical sample purities were above 99\% for the calibration samples and
well above 95\% for the validation samples\cite{thesis:bartsch:2007}.

\begin{figure}[p]
  \vfill\centering
  \includegraphics[width=0.6\textwidth]{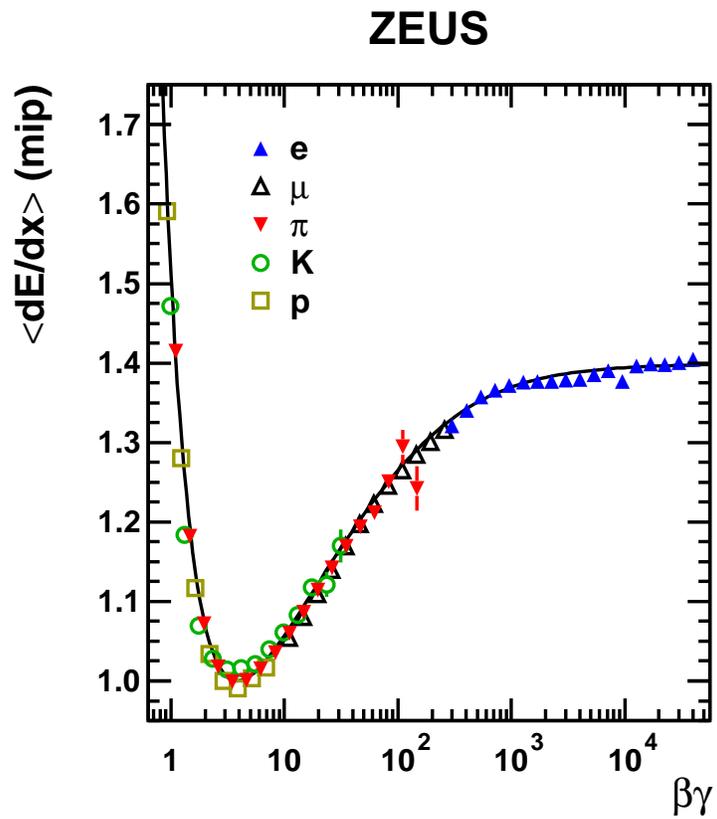}
  \caption{The mean \dEdx{} measured in the CTD, $\langle \dEdx
    \rangle$, as a function of $\beta\gamma$ for different samples of
    identified particles as denoted in the figure. The curve shows a
    physically motivated parametrisation of the $\langle \dEdx \rangle$
    dependence on $\beta\gamma$.}
  \label{fig:dedxcalib}
  \vfill
\end{figure}

After all corrections, the measured \dEdx{} depended only on the
ratio of the particle's momentum to its mass, $\beta\gamma$.
This is illustrated in
Fig.~\ref{fig:dedxcalib}. It shows the specific energy loss as a
function of $\beta\gamma$, for the different samples of identified particles,
$e^{\pm}, \mu^{\pm}, \pi^{\pm}, K^{\pm}, p, \pbar$.
All particle types are well described using a single
physically motivated parametrisation of the mean energy loss as a
function of $\beta\gamma$ with five free parameters following Allison and
Cobb (AC)\cite{allison:80}. 

\begin{figure}[p]
  \vfill\centering
  \includegraphics[width=12cm]{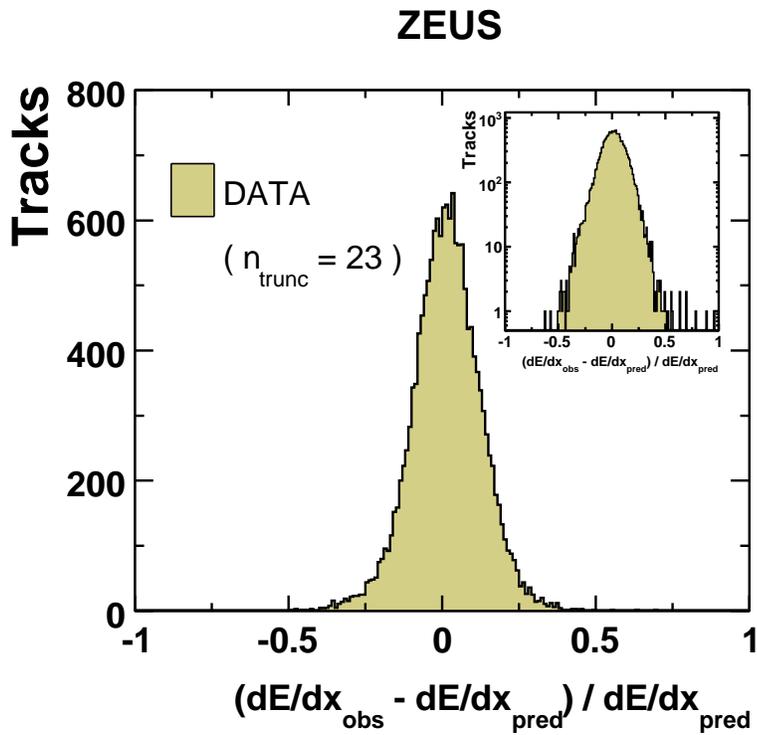}
  \caption{Distribution of the relative difference between the
    observed ($\dEdx_{\mathrm{obs}}$) and predicted
    ($\dEdx_{\mathrm{pred}}$) specific energy loss for the track
    sample with $\ntrunc = 23$. The inset shows the same distribution
    with a logarithmic ordinate scale.  }
  \label{fig:dedxresl}
  \vfill
\end{figure}

Given the quality of the description of the mean \dEdx{} by the AC
parametrisation, the measurements can be used to determine residuals
on \dEdx.  As an example, the distribution of residuals for a sample of
tracks with the number of hits after truncation, \ntrunc, equal to 23
is shown in Fig.~\ref{fig:dedxresl}.  The \dEdx{} resolution is
typically 11\% for tracks that pass at least
five superlayers.  It
improves to about 9\% for tracks that pass all superlayers.

% ----------------------------------------------------------------------------
%       Monte Carlo section
% ----------------------------------------------------------------------------
\section{Monte Carlo simulation}
\label{sec:montecarlo}

To evaluate the detector acceptance and to provide the signal and
background distributions, Monte Carlo samples of beauty, charm and
light-flavour events
generated with
\PYTHIA~6.2\cite{cpc:135:238,*epj:c17:137,*hep-ph-0108264} were used.

The production of \bbbar{}-pairs was simulated following the standard
\PYTHIA{} prescription with the following subprocesses\cite{thesis:kind:2007}:
\begin{itemize}
\item direct and resolved photoproduction with a leading-order massive
  matrix element;
\item $b$ excitation in both the proton and the photon with a
  leading-order massless matrix element.
\end{itemize}

The CTEQ4L\cite{pr:d55:1280} parton
distributions were used for the proton, while GRV-G LO\cite{pr:d46:1973}
was used for the photon.  The $b$-quark mass parameter was set to
$4.75\gev$.
The production of charm and light quarks was simulated for both direct
and non-direct photoproduction with leading-order matrix elements in
the massless scheme
using the same parton distributions as for the \bbbar{} samples.  

The generated events were passed through a full simulation of the ZEUS
detector based on \textsc{Geant}~3.13\cite{tech:cern-dd-ee-84-1}.  The
ionisation loss in the CTD was treated separately using a
parametrisation of the measured data distributions based on the
calibration sample\cite{thesis:kind:2007,thesis:zimmermann:2007}.  The
final Monte Carlo events had to fulfil the same trigger requirements
and pass the same reconstruction programme as the data.

% ----------------------------------------------------------------------------
%       Data selection
% ----------------------------------------------------------------------------
\section{Data selection}
\label{sec:data}

Events were selected online with a
three-level trigger~\cite{zeus:1993:bluebook,proc:chep:1992:222} which required two
jets reconstructed in the calorimeter.

The hadronic system (including the decay electron) was reconstructed
from energy-flow objects (EFOs)\cite{epj:c1:81,*thesis:briskin:1998}
which combine the information from calorimetry and tracking, corrected
for energy loss in inactive material.  Each EFO was assigned a reconstructed four-momentum
$q^{i} = (p_X^i,p_Y^i,p_Z^i,E^{i})$, assuming the pion mass.  Jets
were reconstructed from EFOs using the $k_{T}$
algorithm\cite{pr:d48:3160} in the longitudinally invariant mode with
the massive recombination scheme\cite{np:b406:187} in which $\qjet = \sum_{i}
q^{i}$ and the sum runs over all EFOs.  The transverse energy of
the jet was defined as $\ETjet = \Ejet \cdot \pTjet / \pjet$, where
\Ejet, \pjet{} and \pTjet{} are the energy, momentum and transverse
momentum of the jet, respectively. The transverse energy, \ETjet, is therefore always larger than
the transverse momentum, \pTjet, used in a previous
publication\cite{pr:d70:012008}.

Dijet events were selected as follows:
\begin{itemize}
\item at least two jets with $\ETjet > 7(6)\gev$ for the highest
  (second highest) energetic jet and pseudorapidity of both jets 
  $|\etajet| < 2.5$;
\item the $Z$ coordinate of the reconstructed primary vertex within
  $|Z_{\mathrm{Vtx}}| < 50\cm$;
\item $0.2 < \yJB < 0.8$, where $\yJB =
  (E-P_Z)/(2E_e)$ is the Jacquet-Blondel estimator\cite{proc:epfacility:1979:391}
  for the inelasticity, $y$, and
  $E-P_Z = \sum_i E^i-p_Z^i$, where the sum runs over all EFOs;
\item no scattered-electron
  candidate found in the
  calorimeter with energy $E'_{e} > 5\gev$ and $y_{e} < 0.9$, with
  $y_{e}= 1- \frac{E'_{e}}{2E_{e}}\left( 1-\cos{\theta'_{e}} \right)$, where
  $\theta'_{e}$ is the polar angle of the outgoing
  electron.
\end{itemize}

These cuts suppress background from high-\Qsq{} events and from
non-$ep$ interactions, and correspond to an effective cut of $\Qsq < 1\gev^{2}$.

% ----------------------------------------------------------------------------
%       Identification of electrons from semileptonic decays
% ----------------------------------------------------------------------------
\section{Identification of  electrons from semileptonic decays}
\label{sec:elid}

Electron candidates were selected among the EFOs by
requiring tracks fitted to the primary vertex
and having a transverse momentum, $p_{T}^{e}$,
of at least $0.9\gev$ in the pseudorapidity range $|\eta^{e}| < 1.5$.  
Only the EFOs consisting of a track matched to a single calorimetric cluster
were used.  
To reduce the
hadronic background and improve the overall description, at least $90\%$ of the
EFO energy had to be deposited in the electromagnetic part of the calorimeter.
Electron candidates were required to have a track with $\ntrunc > 12$ to
ensure a reliable \dEdx{} measurement. 
An additional preselection cut of $\dEdx > 1.1\;\mathrm{mip}$ was
applied to reduce the background.
Candidates in the angular region corresponding to the gaps 
between FCAL and BCAL as well as between RCAL and BCAL
were removed using a
cut on the EFO position\cite{thesis:juengst:2005}.

Electrons from photon conversions were tagged and rejected based on
the distance of closest approach of a pair of oppositely charged
tracks to each other in the plane perpendicular to the beam axis and
on their invariant mass~\cite{epj:c18:625}.
Untagged conversion background and electrons from
Dalitz decays were estimated from Monte Carlo studies.

The electron candidate was required to be associated
to a jet using the following procedure:
\begin{itemize}
\item the jet had to have $\ETjet > 6\gev$ and $|\etajet| < 2.5$;
\item the distance $\Delta R = \sqrt{(\etajet-\eta^{e})^{2} + (\phijet
    - \phi^{e})^{2}} < 1.5$;
\item in case of more than one candidate jet, the jet closest in
  $\Delta R$ was chosen.
\end{itemize}

For the identification of electrons from semileptonic heavy-quark
decays, variables for particle identification were combined with
event-based information characteristic of heavy-quark production.  For
a given hypothesis of particle, $i$, and source $j$, the
likelihood, $\mathcal{L}_{ij}$, is given by
\begin{equation*}
  \mathcal{L}_{ij} = \prod\limits_{l}\,\mathcal{P}_{ij}(d_{l})\,,
  \label{eq:partid_likelihood}
\end{equation*}
where $\mathcal{P}_{ij}(d_{l})$ is the probability to observe particle
$i$ from source $j$ with value $d_{l}$ of a discriminant variable.
The particle hypotheses $i \in \{e,\mu,\pi,K,p\}$ and sources, $j$, 
for electrons from semileptonic beauty, charm decays and background,
$j \in \{b,c,\mathrm{Bkg}\}$, were considered.
For the likelihood ratio
test, the test function, $T_{ij}$ was defined as
\begin{equation*}
  T_{ij} = \frac{\alpha_{i} \alpha'_{j} \mathcal{L}_{ij}}{%
    \sum \limits_{m,n} \alpha_{m} \alpha'_{n} \mathcal{L}_{mn}}.
  \label{eq:partid_testfunc}
\end{equation*}
The $\alpha_{i}$, $\alpha'_{j}$ denote the prior probabilities taken
from Monte Carlo.
In the sum, $m,n$ run over all
particle types and sources defined above. 
In the following, $T$ is always taken to be the likelihood ratio for an electron originating from a
semileptonic $b$-quark decay: $T \equiv T_{e,b}$, unless otherwise stated.  
The following five
discriminant variables were combined in the likelihood test:
\begin{itemize}
\item \dEdx, the average energy loss per unit length of the track in the
  CTD;
\item \EMCfrac, the fraction of the EFO energy taken from
  the calorimeter information, \ECAL,
  which is deposited in the electromagnetic part, \EEMC;
\item $E_{\mathrm{CAL}}/p_{\mathrm{track}}$: the EFO energy divided by the track momentum.
\end{itemize}
In order to distinguish
between electrons from semileptonic $b$-quark and $c$-quark decays and
other electron candidates, the following additional observables
were used:
\begin{itemize}
\item \pTrel, the transverse-momentum component of the electron
  candidate relative to the direction of the associated jet defined
  as
  \begin{equation*}
    \label{eq:ptrel}
    \pTrel = \frac{|\vec{p}_{\mathrm{jet}} \times 
      \vec{p}_{e}|}{|\vec{p}_{\mathrm{jet}}|}\,,
  \end{equation*}
  where $\vec{p}_{e}$ is the momentum of the electron candidate.
  The variable \pTrel{} can be used to discriminate between electrons from
  semileptonic $b$-quark decays and from other sources, because its
  distribution depends on the mass of the decaying particle.  It is not
  possible to distinguish charm from light-flavour decays with this variable;
\item \Dphi, the difference of azimuthal angles of the
  electron candidate and the missing transverse momentum vector
  defined as
  \begin{equation*}
    \label{eq:delphi}
    \Dphi = |\phi(\vec{p}_{e})-\phi(\pTmiss)|\,,
  \end{equation*}
  where $\pTmiss$ is the negative vector sum of the EFO momentum transverse to the
  beam axis,
  \begin{equation*}
    \pTmiss = - \textstyle{\left(\sum_{i} p_{x}^{i}, \sum_{i} p_{y}^{i}\right)},
  \end{equation*}
  and the sum runs over all EFOs.
  The vector \pTmiss{} is used as an estimator of the direction of the neutrino from the semileptonic decay.
  The variable \Dphi{} can be used to discriminate
  semileptonic decays of $b$ quarks and $c$ quarks from other sources.

\end{itemize}
 
The shapes of the charm- and light-quark \pTrel{} distributions in the
Monte Carlo were corrected\cite{pr:d70:012008}
using a dedicated background sample in the data.
The value of the correction increased with \pTrel{} and was $15\%$ at
$\pTrel = 1.5\gev$, where the purity of the $b$ contribution is
highest.
For the \Dphi{} distribution
a correction was determined in a similar way, but in this
case the maximal correction was only of the order of $5\%$.  

\begin{figure}[p]
  \vfill\centering
  \includegraphics[bb = 0 0 475 545,clip=true,width=\textwidth]{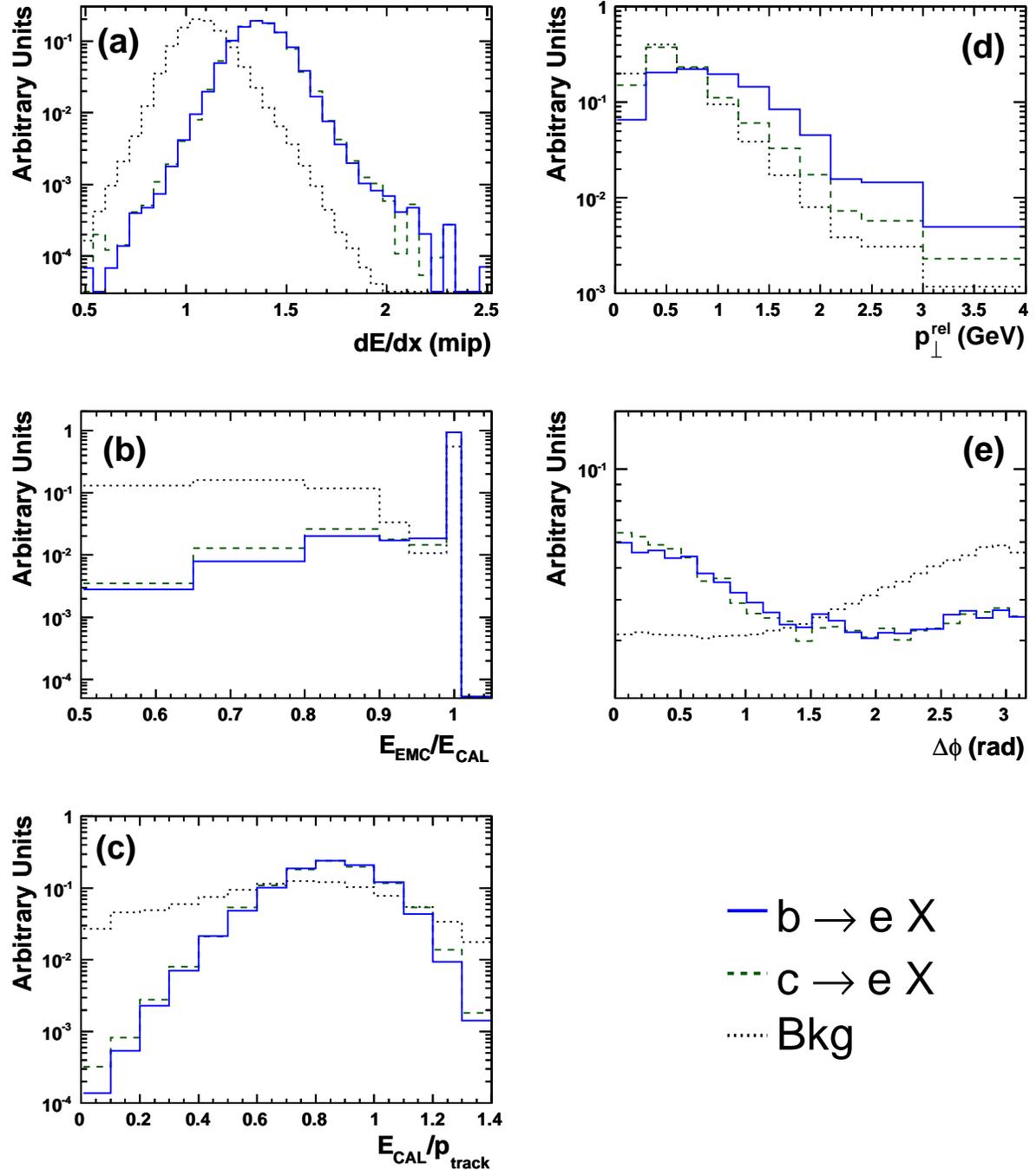}
  \caption{Normalised distributions of the five input variables used in the likelihood 
    for the electron candidates extracted from the Monte Carlo,
    without applying the cuts on \dEdx{} and \EMCfrac.
    The solid line shows the distribution for electrons from semileptonic
    $b$-quark decays, the dashed line for $c$-quark decays and the dotted line
    the background (Bkg).}
  \label{fig:likel_pdf}
  \vfill
\end{figure}

In Fig.~\ref{fig:likel_pdf} the distributions of the five input
variables used in the likelihood
are shown for electrons from $b$-quark and $c$-quark decays and for
electron candidates from other sources.
A clear
difference in shape between signal and background can be seen. 

% ----------------------------------------------------------------------------
%       Signal extraction
% ----------------------------------------------------------------------------
\section{Signal extraction}
\label{sec:signal}

\begin{figure}[p]
  \vfill\centering
  \includegraphics[width=\textwidth]{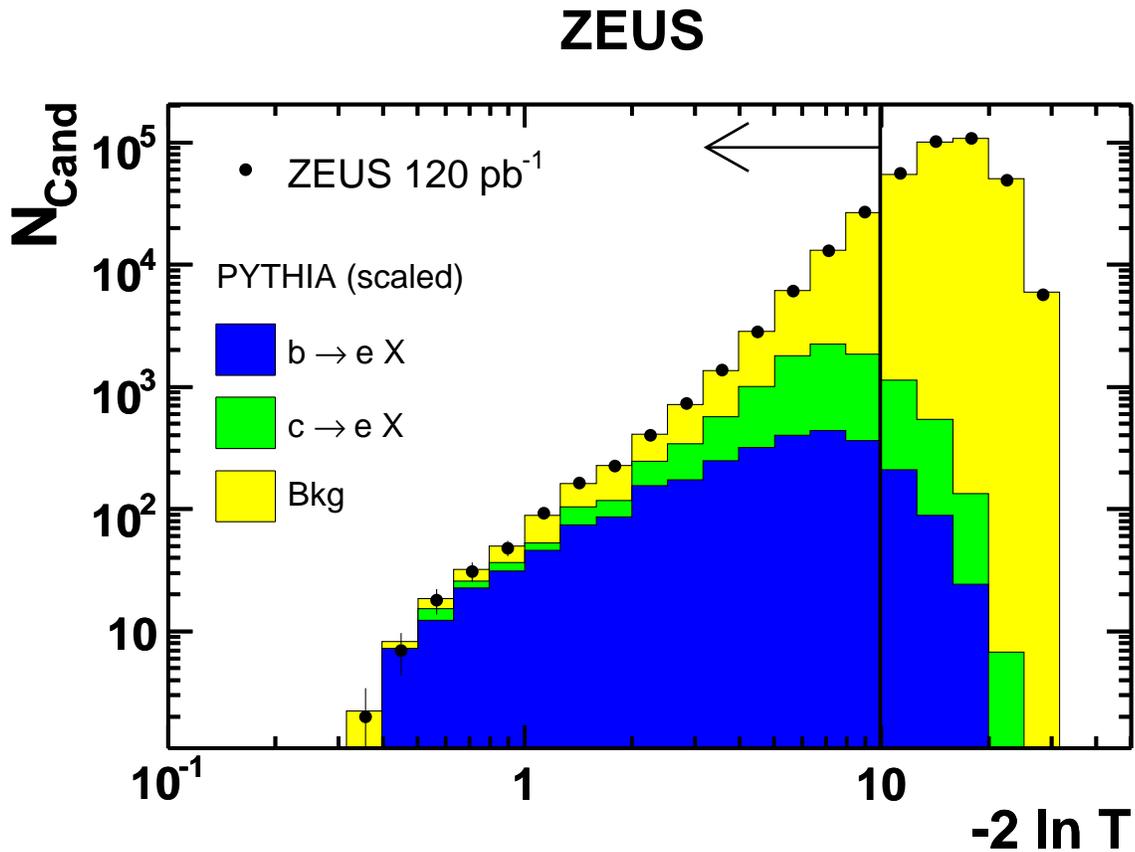}
  \caption{The distribution of the likelihood ratio for electron candidates, \Ncand, in data
    compared to the Monte Carlo expectation after the fit described in the text.
    The arrow indicates the region included in the
    fit ($-2 \ln T < 10)$.
    The shaded areas show the fitted contributions from $b$ quarks,
    $c$ quarks and background as denoted in the figure.}
  \label{fig:lik}
  \vfill
\end{figure}

\begin{figure}[p]
  \vfill\centering
  \includegraphics[bb = 0 0 480 600,clip=true,width=\textwidth]{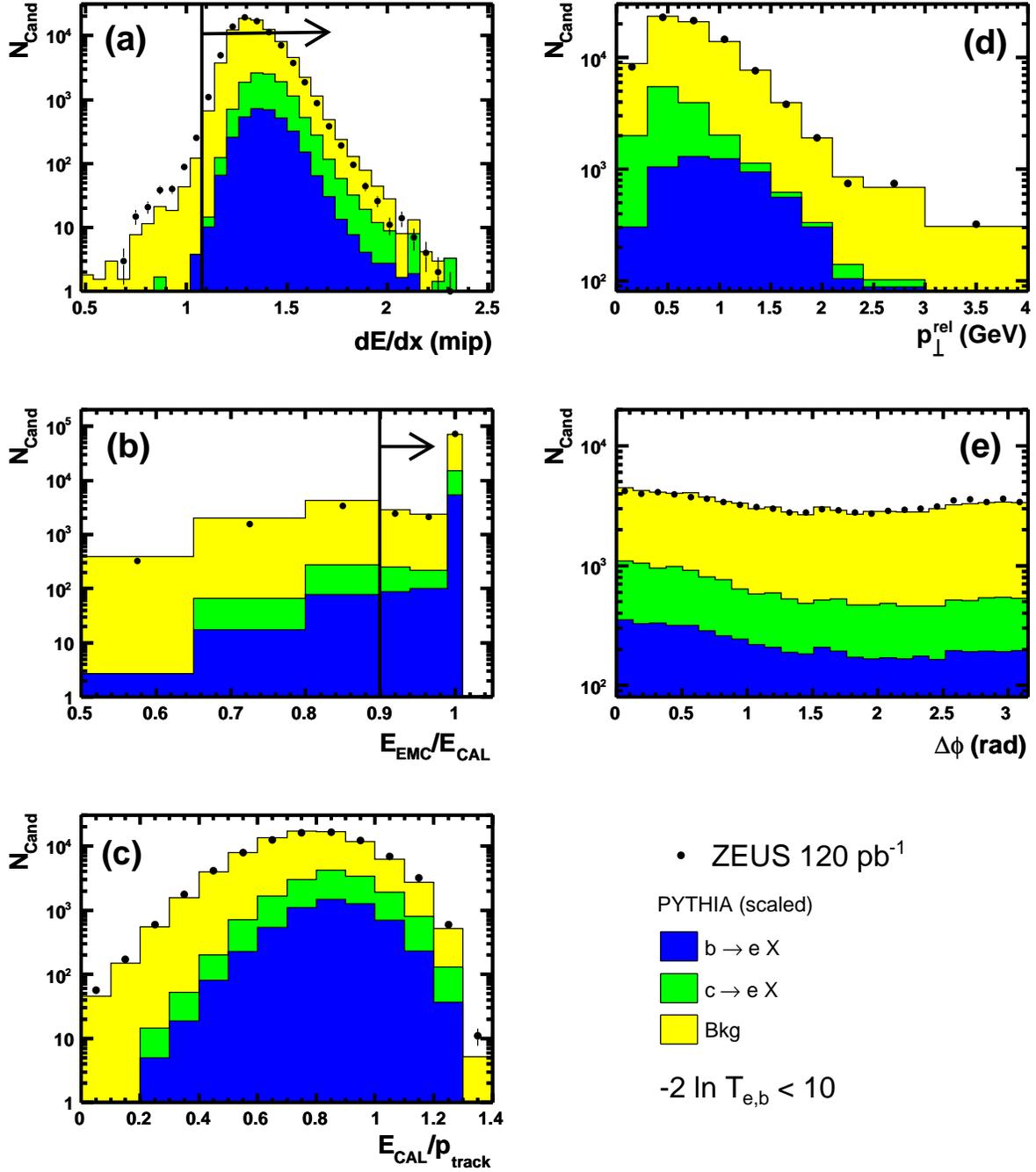}
  \caption{Distributions of the five input variables of the likelihood
    for the electron candidates used in the fit ($-2 \ln T < 10$).
    All cuts have been applied except $\dEdx > 1.1$ in a) and
    $\EMCfrac > 0.9$ in b) (the cuts are indicated in the figure).
    The shaded areas show the contributions from $b$ quarks, $c$
    quarks and background as denoted in the figure, after applying the
    scale factors from the fit.  }
    \label{fig:ctrl_likel_all}
    \vfill
\end{figure}

\begin{figure}[p]
  \vfill\centering
  \includegraphics[bb = 0 0 480 600,clip=true,width=\textwidth]{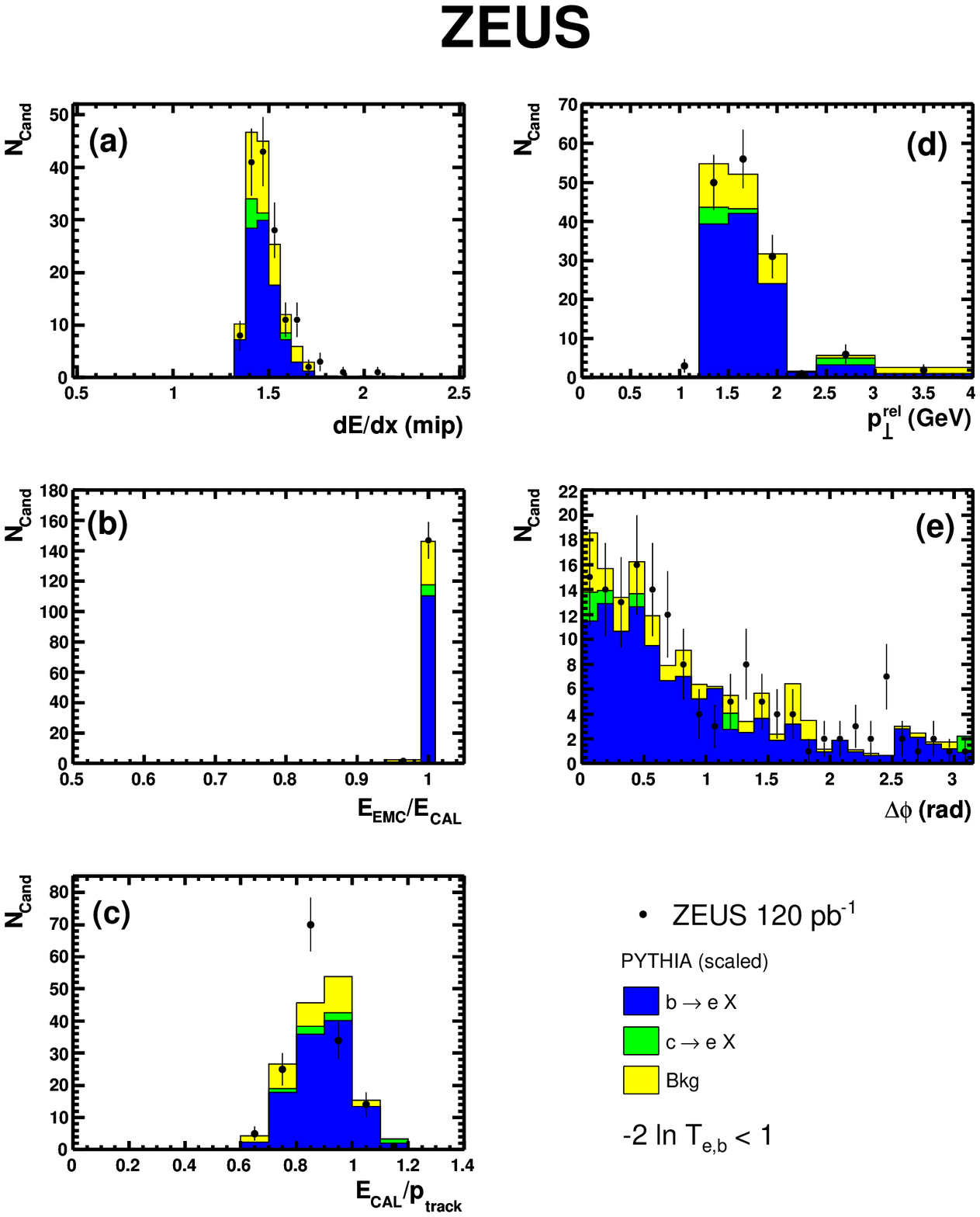}
  \caption{Distributions of the five input variables of the likelihood 
    for a beauty enriched sample (candidates with $-2 \ln T < 1$). 
    In these plots the ordinate is shown on a linear scale.
    Other details as in the caption of Fig.~\protect\ref{fig:ctrl_likel_all}.
  }
  \label{fig:ctrl_likel_beauty}
  \vfill
\end{figure}

\begin{figure}[p]
  \vfill\centering
  \includegraphics[bb = 0 0 480 600,clip=true,width=\textwidth]{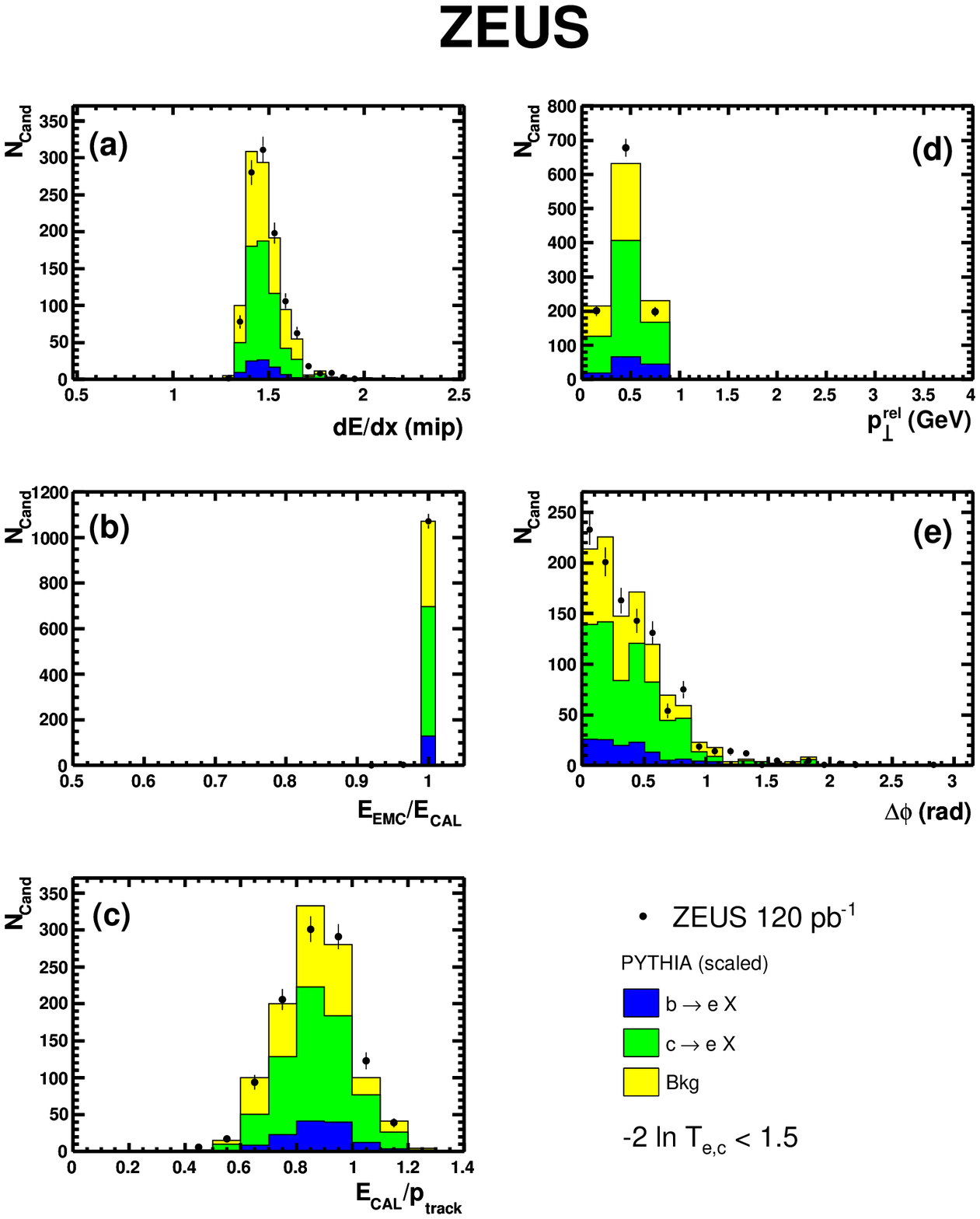}
  \caption{Distributions of the five input variables of the likelihood 
    for a charm enriched sample (candidates with $-2 \ln T_{e,c} < 1.5$).  
    In these plots the ordinate is shown on a linear scale.
    Other details as in the caption of Fig.~\protect\ref{fig:ctrl_likel_all}.
  }
  \label{fig:ctrl_likel_charm}
  \vfill
\end{figure}

\begin{figure}[p]
  \vfill\centering
  \includegraphics[bb=0 0 560 535, width=\textwidth]{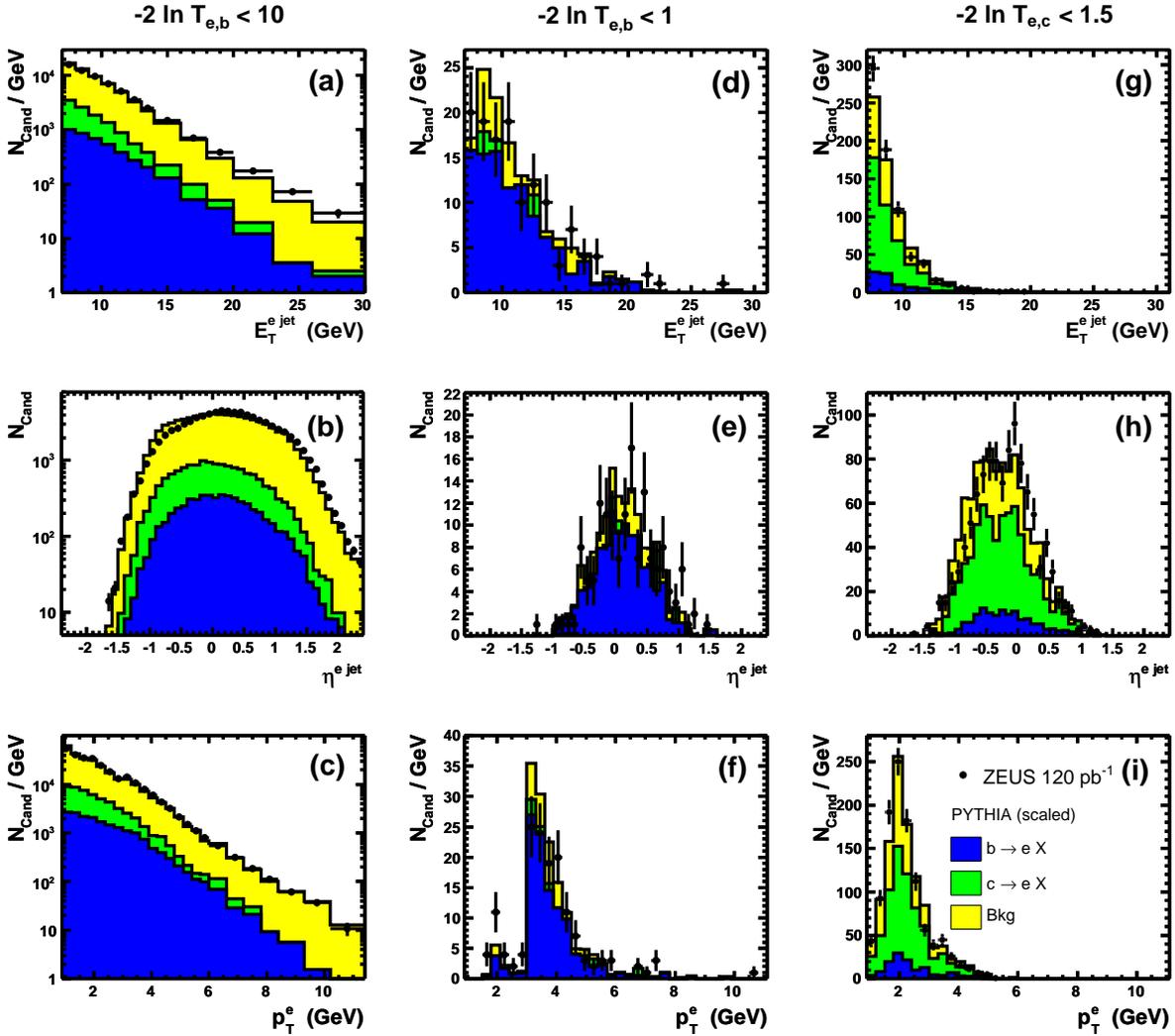}
  \caption{Distributions of \ETejet{} and \etaejet{} of the jet associated
    with the electron, and \pTe{} of the electron candidate. Figures
    a)-c) contain all electron candidates satisfying $-2 \ln T_{e,b} < 10$;
    d)-f) and g)-i) show the same distributions for the beauty ($-2 \ln T_{e,b} < 1$)
    and charm ($-2 \ln T_{e,c} < 1.5$) enriched samples,
    respectively.
    Other details as in the caption of Fig.~\protect\ref{fig:ctrl_likel_all}.
    }
  \label{fig:control}
  \vfill
\end{figure}

The electron candidates in the Monte Carlo samples
were classified as originating from beauty, charm or background. 
The beauty sample also contains the cascade decays $b \rightarrow c
\rightarrow e$, but not $b \rightarrow \tau \rightarrow e$ and $b
\rightarrow \Jpsi \rightarrow e^{+}e^{-}$ that have been included in
the background sample. Test functions (see Section~\ref{sec:elid})
were calculated separately for the three samples. The fractions of the
three samples in the data, $f_{e,b}^{\mathrm{DATA}}$,
$f_{e,c}^{\mathrm{DATA}}$, $f_{\mathrm{Bkg}}^{\mathrm{DATA}}$, were
obtained from a three-component maximum-likelihood fit\cite{barlow:93}
to the $T$ distributions. The constraint $f_{e,b}^{\mathrm{DATA}}
+f_{e,c}^{\mathrm{DATA}}+ f_{\mathrm{Bkg}}^{\mathrm{DATA}} = 1$ was
imposed in the fit.  The fit range of the test function was restricted
to $-2 \ln T < 10$ to remove the region dominated by background and
where the test function falls rapidly.  The $\chi^{2}$ for the fit is
$\chi^{2}/\mathrm{ndf} = 13/12$ and the $b$-quark and $c$-quark
measurements have a correlation coefficient of $-0.6$.  The result of
the fit is shown in Fig.~\ref{fig:lik} and corresponds to a scaling of
the cross section predicted by the beauty Monte Carlo by a factor of
$1.75 \pm 0.16$ and the charm Monte Carlo by a factor of $1.28 \pm
0.13$. These factors are applied to
Figs.~\ref{fig:ctrl_likel_all}--\ref{fig:control} and denoted as
``PYTHIA~(scaled)''.
A fit over the whole $T$ range gave
consistent cross sections and was used as a
cross-check.

The distributions of the five variables that entered the likelihood are shown
in Fig.~\ref{fig:ctrl_likel_all}. The description of all variables is
reasonable. 
These distributions are dominated by the background contribution. In order to select a
beauty-enriched sample, a cut of $-2 \ln T < 1$ was applied. The
resulting distributions are shown in Fig.~\ref{fig:ctrl_likel_beauty}. A
likelihood for semileptonic charm can also be constructed, $T_{e,c}$. The
distributions of the likelihood for a sample satisfying $-2 \ln T_{e,c} < 1.5$
are shown in Fig.~\ref{fig:ctrl_likel_charm}. Good agreement is
observed in both cases.

To demonstrate the quality of the data description by the Monte Carlo,
the distributions of \ETjet{} and \etajet{} of the
jet associated with the electron and of the \pTe{} of the electron
candidates are compared in Fig.~\ref{fig:control}a)--c).  
In Fig.~\ref{fig:control}d)--i) the same distributions are
compared for the beauty- and charm enriched-samples.
Some differences are observed in the jet variables, mainly in the region
dominated by background. The agreement significantly improves for
samples enriched in beauty and charm signal.

% ----------------------------------------------------------------------------
%       Cross section determination
% ----------------------------------------------------------------------------
\section{Cross section determination}
\label{sec:xsect}

The cross sections have been measured in the kinematic range $\Qsq <
1\gev^{2}$, $0.2 < y < 0.8$, with at least two jets with $\ETjet > 7 (6)\gev$,
$|\etajet| < 2.5$ and an electron from a
semileptonic decay with $\pTe > 0.9\gev$ in the range $|\etae| < 1.5$.

The differential beauty cross section for a variable, $v$, was determined
separately for each bin, $k$, from the relative fractions in the data
obtained from the fit and the acceptance correction,
$\mathcal{A}_{v_{k}}^{b}$, calculated using the Monte
Carlo,
\begin{equation*}
  \label{eq:diffacceptance}
  \frac{d\sigma_{b}}{d v_{k}} = 
  \frac{N^{\mathrm{DATA}} \cdot f_{e,b}^{\mathrm{DATA}}(v_{k})}{%
    \mathcal{A}_{v_{k}}^{b}
    \cdot \mathcal{L} \cdot \Delta v_{k}},
\end{equation*}
where $N^{\mathrm{DATA}}$ is the number of electron candidates found in the
data, $\mathcal{L}$ is the integrated luminosity
and $\Delta v_{k}$ is the bin width.

In order to determine the acceptance, the jet-finding algorithm was
applied to the MC events
after the detector simulation and at hadron level.
The acceptance is defined as
\begin{equation*}
  \mathcal{A} = \frac{N_{e}^{\textrm{obs}}}{N_{e}^{\mathrm{had}}} ,
\end{equation*}
where $N_{e}^{\textrm{obs}}$ is the number of electrons from semileptonic
decays reconstructed in the Monte Carlo
satisfying the selection criteria detailed in Sections~\ref{sec:data}
and \ref{sec:elid}, and
$N_{e}^{\mathrm{had}}$ is the number of
electrons from semileptonic decays produced in the signal process that satisfy
the kinematic requirements using the Monte Carlo information at the
generator level.
At hadron level, the $k_{T}$ algorithm was applied to all final-state particles with 
a lifetime of $\tau > 0.01\ns$ and the electron was associated to its parent jet using the
generator information.

All cross sections were measured separately for the
two centre-of-mass energies $\sqrt{s}=300$ and $318\gev$. Additionally, the
cross sections were calculated with the whole data set and were corrected to
$\sqrt{s}=318\gev$.  The correction factor of $\approx 2\%$ was determined with LO as well
as NLO calculations.

The charm cross sections were measured using the
same procedure.

% ----------------------------------------------------------------------------
%        Systematics and consistency checks
% ----------------------------------------------------------------------------
\section{Systematic uncertainties}
\label{sec:syst}

The systematic uncertainties were calculated by varying the analysis
procedure and then redoing the fit to the likelihood distributions.
The following sources were the main contributors to the systematic
uncertainty (the first value in parentheses is the uncertainty for
beauty, while the second is that for charm):
\begin{itemize}
\item the systematic uncertainty on the description of the \dEdx{}
  information was estimated by looking at the differences between the
  various calibration and validation samples. Variations in the
  mean, width and shape of the distributions were evaluated and used
  as a measure of the uncertainty\cite{thesis:bartsch:2007}. The resulting
  uncertainty was found to be ($^{+1}_{-5}\%$ / $^{+10}_{-3}\%$);

\item the changes in the correction to the \pTrel{} distribution in
  various kinematic ranges were taken as a measure of its
  uncertainty. For $\pTrel = 1.5\gev$ the variation was $20\%$
  of the correction. The changes led to a systematic uncertainty of
   ($^{+3}_{-6} \%$ / $^{+10}_{-5}\%$).

  In addition, the correction to the charm
  distribution was varied from zero to that of the background
  sample. This led to an uncertainty of ($^{+6}_{-4}\%$ / $^{+7}_{-1}\%$);

\item a shift of the CAL energy scale in the Monte Carlo simulation by
  $\pm 3 \%$ ($\pm 2\%$ / $\pm 5\%$);

\item reweighting of the direct and non-direct contributions in the
  Monte Carlo to provide a better description of the data ($+1\%$ /
  $+3\%$);

\item the estimated residual number of electrons left in the sample
  from photon conversions as well
  as from Dalitz decays were varied by 25\% and 20\%
  respectively\cite{thesis:verkerke:1998}.  This led to systematic
  uncertainties of ($\pm 1\%$ / $\mp 4\%$) due to photon conversions and
  ($\pm 1\%$ / $\mp 1\%$) due to Dalitz decays.
\end{itemize}

These systematic uncertainties were added in quadrature separately for the negative
and the positive variations to determine the overall systematic uncertainty of
$^{+8}_{-9}\%$ for the beauty and $^{+17}_{-9}\%$ for the
charm cross sections. Since no significant dependence of the
systematic uncertainties on the kinematic
variables was observed, the same
uncertainty was applied to each data point.
A 2\% overall normalisation uncertainty associated with the luminosity
measurement was included.

A series of further checks were made.  The cut on the transverse
momentum of the electron candidate was varied by $\pm 3\%$,
which is the momentum uncertainty for a track with $\pT =
0.9\gev$.  The \Dphi{} correction was varied within its
uncertainty. 
The cut on
$\Delta R$ to associate the decay electron with a jet was varied between
1.5 and 1.0. 
The effect of the gaps between FCAL and BCAL as well as between RCAL and BCAL
was investigated by varying the cut on the EFO position.
Various tests of the signal-extraction method were made: e.g.\ using the
likelihood without the \EMCfrac{} or
$E_{\mathrm{CAL}}/p_{\mathrm{track}}$ variables; applying the fit on a
signal-enriched sample by making tighter cuts on the input
variables and varying the fit range.
The prior
probabilities were recalculated after the fit and used as the input for a
second fit iteration. 
Separate fits were made for electron and positron candidates for each of the
lepton beam particles ($e^{-}$ and $e^{+}$) separately as well as for the
combined sample.  
All variations were found to be
consistent with the expected fluctuations due to statistics; therefore
they have not been added to the systematic uncertainty.

% ----------------------------------------------------------------------------
%       Theory
% ----------------------------------------------------------------------------
\section{Theoretical predictions and uncertainties}
\label{sec:theory}

QCD predictions at NLO, based on the FMNR
programme\cite{np:b412:225,*asdhep:15:609}, are compared to the
data. The programme separately generates processes containing point-like and
hadron-like photon contributions, which have to be
combined to obtain the total cross section.  
The $b\bbar$
and the $c\cbar$ production cross sections were calculated separately.
The parton distribution 
functions were taken from CTEQ5M\cite{epj:c12:375} for the proton and
GRV-G HO\cite{pr:d46:1973} for the photon.  
The heavy-quark masses (pole masses) were set to $m_{b}=4.75\gev$ and
$m_{c}=1.6\gev$. 
The strong coupling constant, $\Lambda_{\mathrm{QCD}}^{(5)}$, was set to
0.226\gev.  The renormalisation, $\mu_{R}$, and factorisation,
$\mu_{F}$, scales were chosen to be equal and set to
$\mu_{R}=\mu_{F}=\sqrt{\hat{p}_{T}^{2}+m_{b(c)}^{2}}$, where
$\hat{p}_{T}$ is the average transverse momentum of the heavy quarks.

The Peterson fragmentation function\cite{pr:d27:105}, with
$\epsilon_{b}=0.0035$ and $\epsilon_{c}=0.035$\cite{np:b565:245}, was
used to produce beauty and charm hadrons from the heavy quarks.  For
the $b\bbar$ and $c\cbar$ cross sections, the decays into electrons
were simulated using decay spectra from \PYTHIA.  

For beauty, both the contributions from prompt and from
cascade decays, excluding $b \rightarrow \tau \rightarrow e$ and $b
\rightarrow \Jpsi \rightarrow e^{+}e^{-}$, are taken into account in
the effective branching fraction. The values were set to 0.221 for the
\bbbar{} and to 0.096 for the \ccbar{} cross
sections\cite{jphys:g33:1}.

For the systematic uncertainty on the theoretical prediction, the
masses and scales were varied simultaneously to maximise the change in
the cross section using the values: $m_{b} = 4.5,5.0\gev$, $m_{c} =
1.35,1.85\gev$ and
$\mu_{R}=\mu_{F}=\frac{1}{2}\sqrt{\hat{p}_{T}^{2}+m_{b(c)}^{2}}, 2
\sqrt{\hat{p}_{T}^{2}+m_{b(c)}^{2}}$.  The effects of different parton
density functions as well as variations of $\epsilon_{b}$ within the
uncertainty of $0.0015$ had a small effect on the cross-section
predictions and were neglected.  The parameter $\epsilon_{c}$ was
varied between $0.02$ and $0.07$ and the contribution was added in
quadrature to the systematic uncertainty.  The uncertainty on the
electron decay spectra, evaluated from comparisons to experimental
measurements~\cite{pl:b547:181,*pr:d67:031101,prl:97:251801} and to a
simple free-quark decay model, was found to be small compared to the
total theoretical uncertainty and was neglected.

The uncertainty on the NLO QCD predictions for the total cross section
are $+25\%$ and $-15\%$ for beauty and $+45\%$ and $-28\%$ for charm.

The NLO QCD predictions for parton-level jets, reconstructed by
applying the $k_{T}$ algorithm to the outgoing partons, were corrected for
hadronisation effects. A bin-by-bin procedure was used according to
$d\sigma = d\sigma_{\mathrm{NLO}} \cdot \Chad$, where
$d\sigma_{\mathrm{NLO}}$ is the cross section for partons in the final
state of the NLO calculation. The hadronisation correction factor,
\Chad, was defined as the ratio of the dijet cross sections, extracted
from the \PYTHIA{} Monte Carlo, after and
before the hadronisation process, $\Chad =
d\sigma_{\mathrm{MC}}^{\mathrm{Hadrons}}
/d\sigma_{\mathrm{MC}}^{\mathrm{Partons}}$.  The hadronic corrections
were generally small and are given in
Tables~\ref{tab:visxsection}--\ref{tab:jetb}. No uncertainty was assigned
to the correction.

% ----------------------------------------------------------------------------
%       Results
% ----------------------------------------------------------------------------
\section{Results}
\label{sec:results}

\begin{figure}[p]
  \vfill\centering
  \includegraphics[width=10cm]{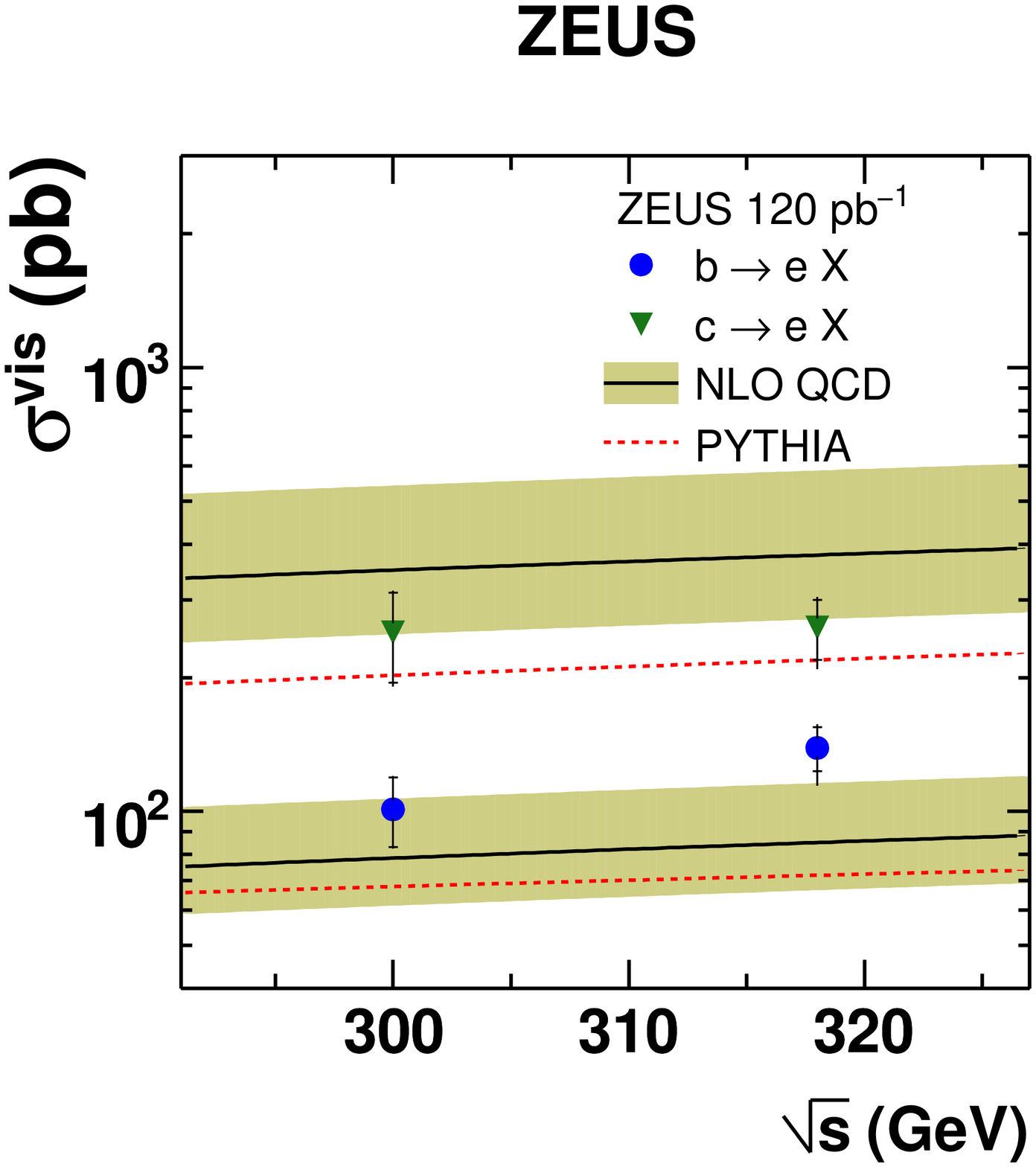}
  \caption{Total cross sections for electrons 
    from $b$ and $c$ quarks in photoproduction events, 
    $\Qsq < 1\gev^{2}$ and $0.2 < y < 0.8$, with 
    at least two jets with $\ET > 7(6)\gev$, $|\eta| < 2.5$ 
    and the subsequent semileptonic decay to an electron with $\pT >
    0.9\gev$ and $|\eta| < 1.5$.
    The measurements are shown as points.
    The inner error bar shows the statistical uncertainty and the
    outer error bar shows
    the statistical and systematic uncertainties added in quadrature.
    The solid line shows the NLO QCD prediction after hadronisation corrections, 
    with the theoretical uncertainties indicated by the band; the
    dashed line shows the prediction from \PYTHIA.}
  \label{fig:totxsect}
  \vfill
\end{figure}

The visible $ep$ cross sections (quoted at hadron level) for $b$-quark and
$c$-quark production and the subsequent semileptonic decay to an
electron with $\pTe > 0.9\gev$ in the range $|\etae| < 1.5$ in
photoproduction events with $\Qsq < 1 \gev^{2}$ and $0.2 < y < 0.8$ and at
least two jets with $\ET > 7 (6)\gev$, $|\eta| < 2.5$ were
determined separately for $\sqrt{s}=300\gev$ and
$\sqrt{s}=318\gev$. The measurements are given in
Table~\ref{tab:visxsection} and are shown in Fig.~\ref{fig:totxsect}.
The cross sections at the two different centre-of-mass energies are
consistent with each other; combining the results leads to a reduced
statistical uncertainty.  
For
the complete data set ($96 \rnge 00$) the cross sections evaluated at
$\sqrt{s}=318\gev$ are
\begin{align*}
\sigvisb & = 
\left( 125 \pm 11 (\stat) ^{+10}_{-11} (\syst) \right)\pb,\\ 
\sigvisc & = 
\left( 278 \pm 33 (\stat) ^{+48}_{-24} (\syst) \right)\pb.
\end{align*}

The visible $b$-quark cross section was also determined in the kinematic region
of a previous ZEUS analysis using muons\cite{pr:d70:012008} and is
in good agreement.

The NLO QCD predictions of FMNR (see Table~\ref{tab:visxsection}) are
compared to the data in Fig.~\ref{fig:totxsect}.
Good agreement is observed.
Also shown in Fig.~\ref{fig:totxsect} are expectations of the \PYTHIA{} Monte Carlo.
The combined \bpbbar{} cross section is a factor 1.75 higher
while the \cpcbar{} cross section is a factor of 1.28 higher than the \PYTHIA{}
prediction (see Section~\ref{sec:signal}). These factors
are used to scale the \PYTHIA{} predictions in the following figures.

\begin{figure}[p]
  \vfill\centering
  \includegraphics[bb = 0 0 500 530, clip=true,width=\textwidth]{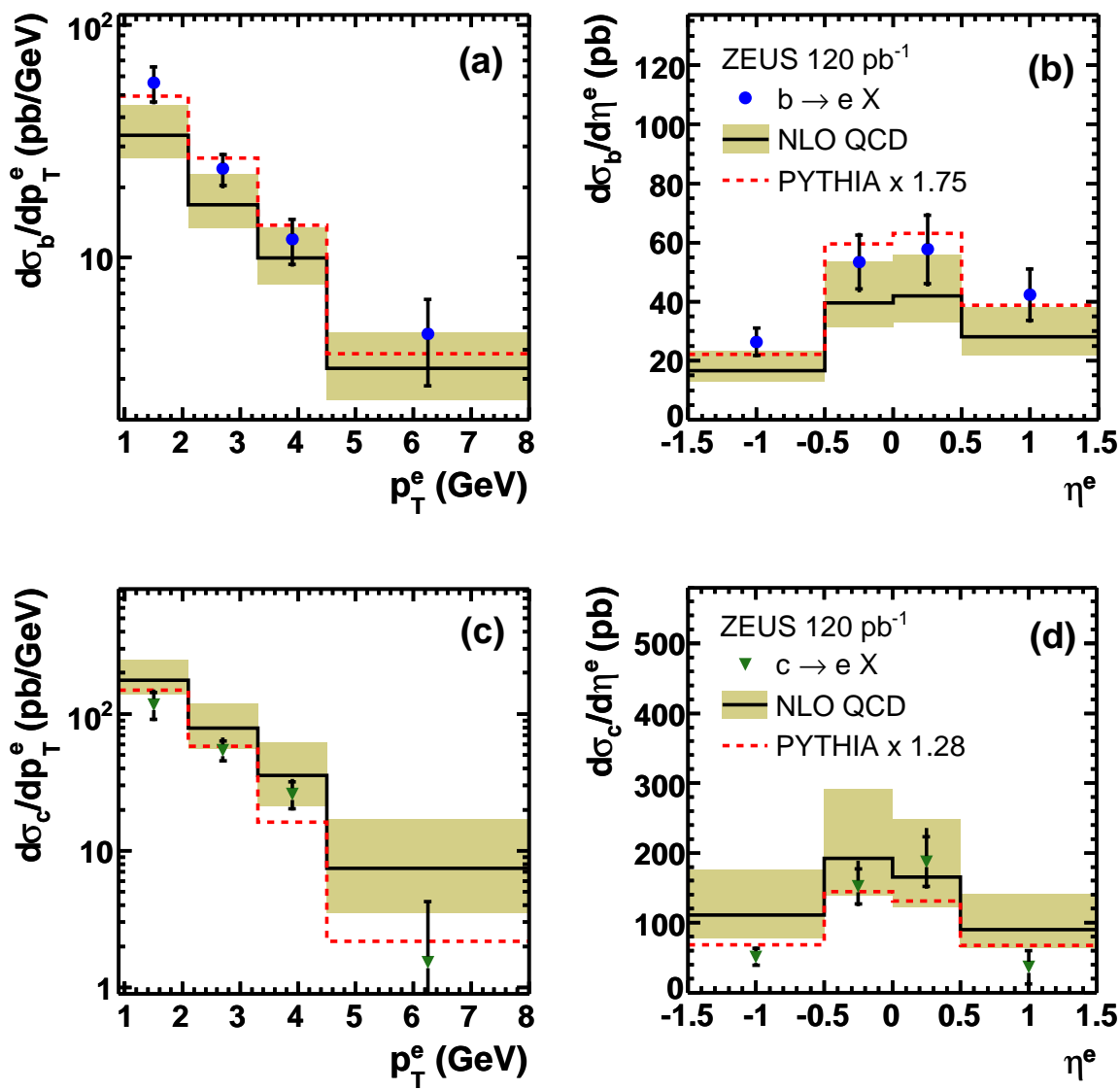}
  \caption{Differential cross sections  as a function of 
    a), c) the transverse momentum
    and b), d) the pseudorapidity of the electrons. Plots a) and b) are for
    $b$-quark production while c) and d) are for $c$-quark
    production.
    The measurements are shown as points. The inner error bar shows the
    statistical uncertainty and the outer error bar shows the statistical and
    systematic uncertainties added in quadrature.
    The solid line shows the NLO QCD prediction after hadronisation corrections, 
    with the theoretical uncertainties indicated by the band; the
    dashed line shows the scaled prediction from \PYTHIA.
    }
  \label{fig:diffxsect-elec}
  \vfill
\end{figure}

\begin{figure}[p]
  \vfill\centering
  \includegraphics[bb = 0 0 250 530,clip=true,width=0.5\textwidth]{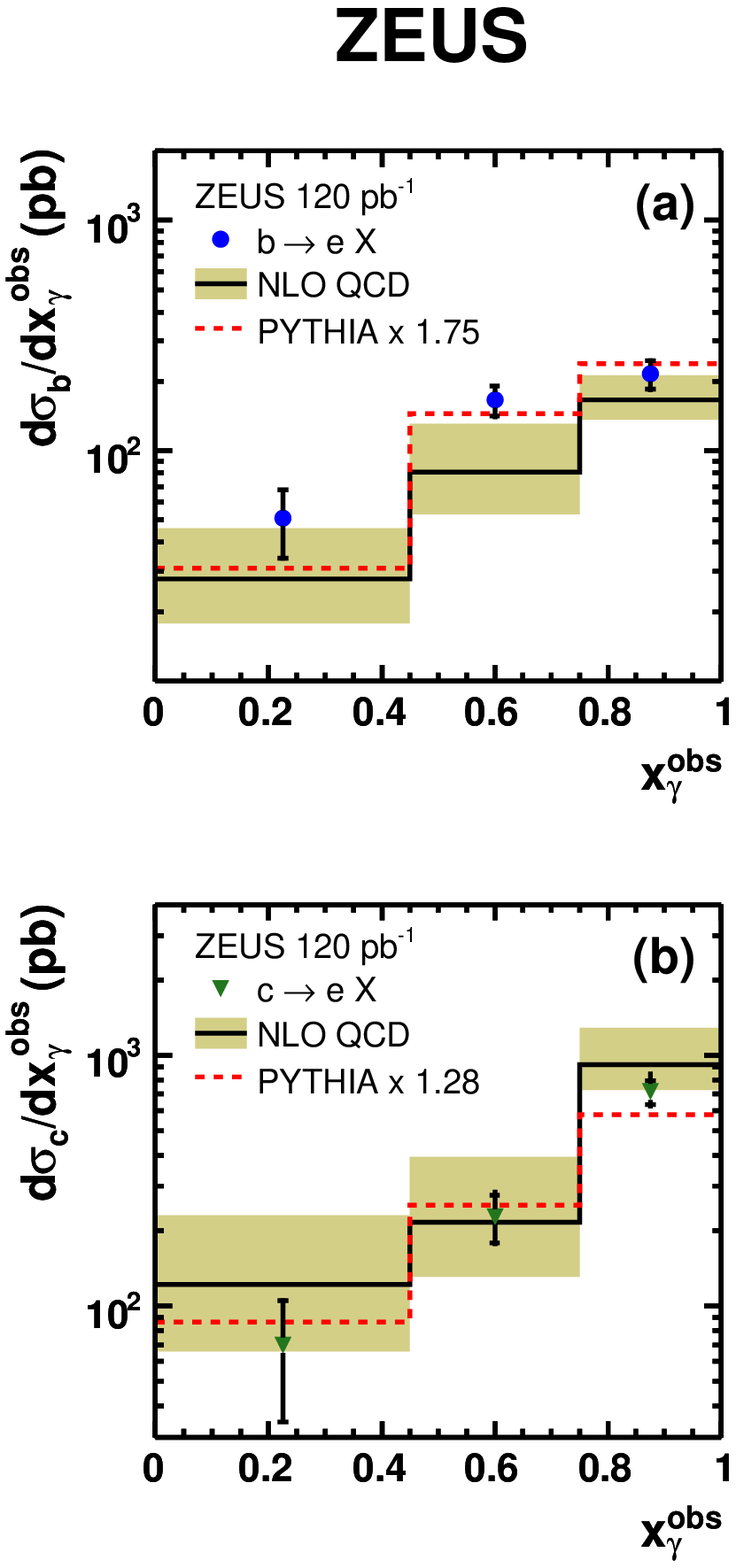}
  \caption{Differential cross sections  as a function of \xgamobs.
    a) shows the distribution for electrons from $b$-quark production while 
    b) shows $c$-quark production.
    Other details as in the caption of Fig.~\protect\ref{fig:diffxsect-elec}.
  }
  \label{fig:diffxsect-xgamma}
  \vfill
\end{figure}

\begin{figure}[p]
  \vfill\centering
  \includegraphics[bb = 0 0 500 530, clip=true, width=\textwidth]{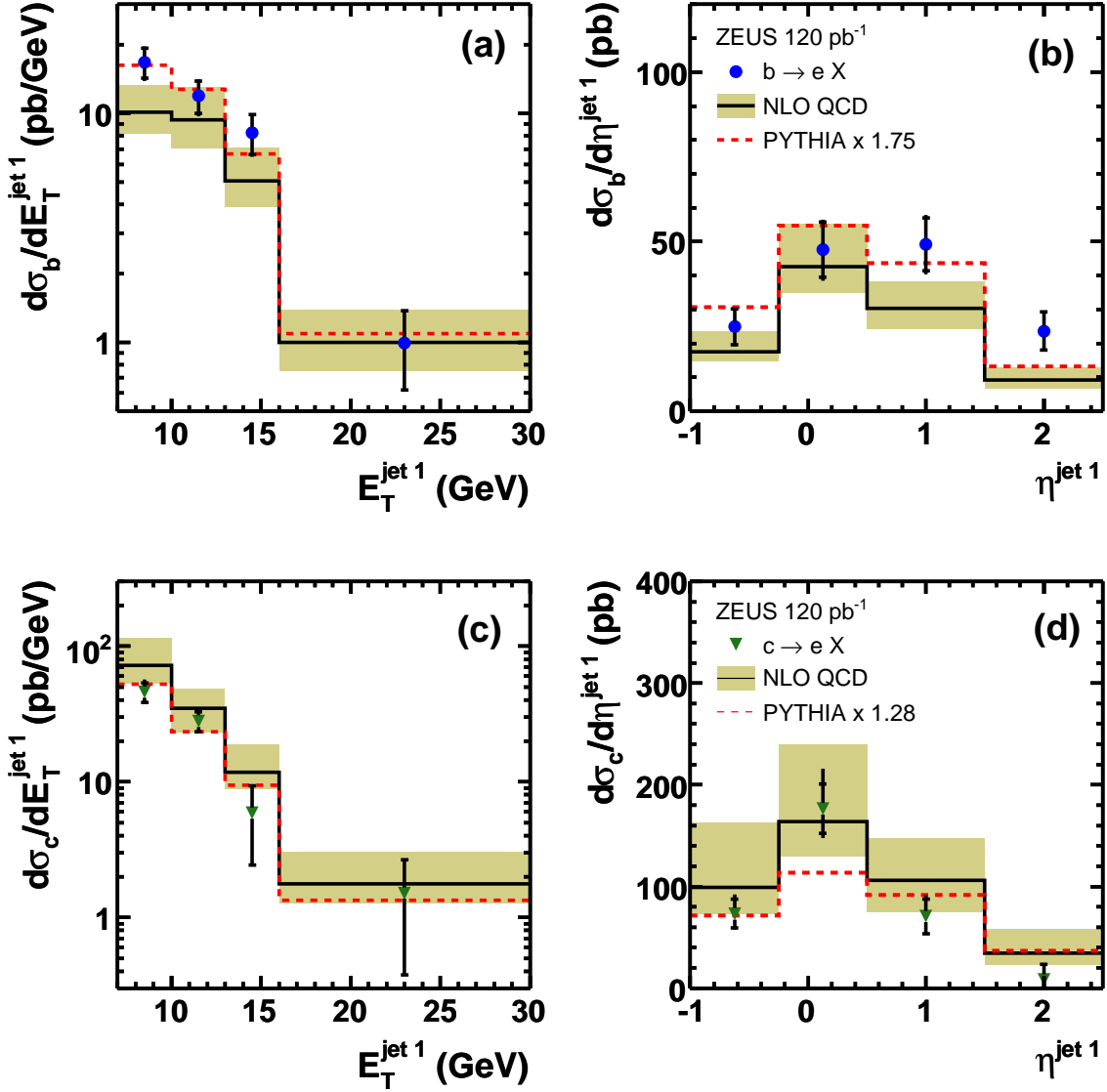}
  \caption{Differential cross sections as a function of a), c) the
    transverse energy and b), d) the pseudorapidity of the highest-energy jet. Plots a) \& b)
    show the distributions for electrons from $b$-quark production
    while plots c) \& d) 
    show those for $c$-quark production.
    Other details as in the caption of Fig.~\protect\ref{fig:diffxsect-elec}.
  }
 
  \label{fig:diffxsect-jet}
  \vfill
\end{figure}

Differential cross sections as a function of
\pTe{} and $\eta_{e}$, \xgamobs, \ETjeti{} and \etajeti{} are shown in
Figs.~\ref{fig:diffxsect-elec}, \ref{fig:diffxsect-xgamma} and
\ref{fig:diffxsect-jet}, respectively.  
The variable \xgamobs{} is defined as
\begin{equation*}
  \xgamobs = \frac{\sum_{i=1,2} (E^{\mathrm{jet}\,i} - p^{\mathrm{jet}\,i}_Z)}{E-p_Z},
\end{equation*}
where the sum is over the two highest-energy jets, and
corresponds at leading order to the fraction of the
exchanged-photon momentum in the hard scattering process.
The figures also
show the NLO QCD and the scaled \PYTHIA{} predictions.
The cross-section values are
given in Tables~\ref{tab:electron}--\ref{tab:jet1}.  
Both the predictions from the NLO QCD calculations as well as the scaled
\PYTHIA{} cross sections describe the data well.

\begin{figure}[p]
  \vfill\centering
  \includegraphics[bb=0 0 250 530, clip=true,width=0.5\textwidth]{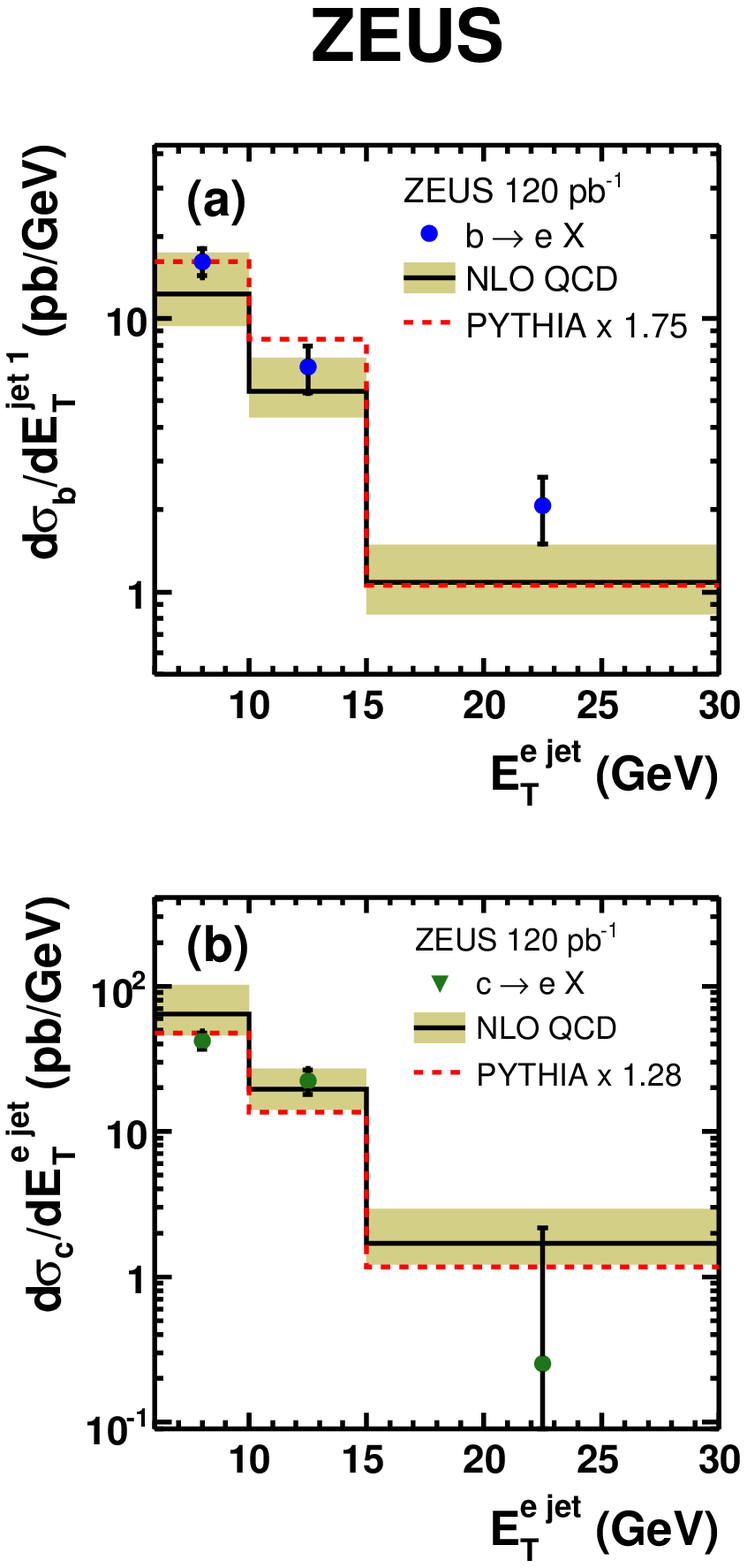}
  \caption{Differential cross sections for a) $b$-quark and b) $c$-quark
    production as a function of the transverse energy of the jet
    associated to the electron.
    Other details as in the caption of Fig.~\protect\ref{fig:diffxsect-elec}.
    }
  \label{fig:etb}
  \vfill
\end{figure}

\begin{figure}[p]
  \vfill\centering
  \includegraphics[width=\textwidth]{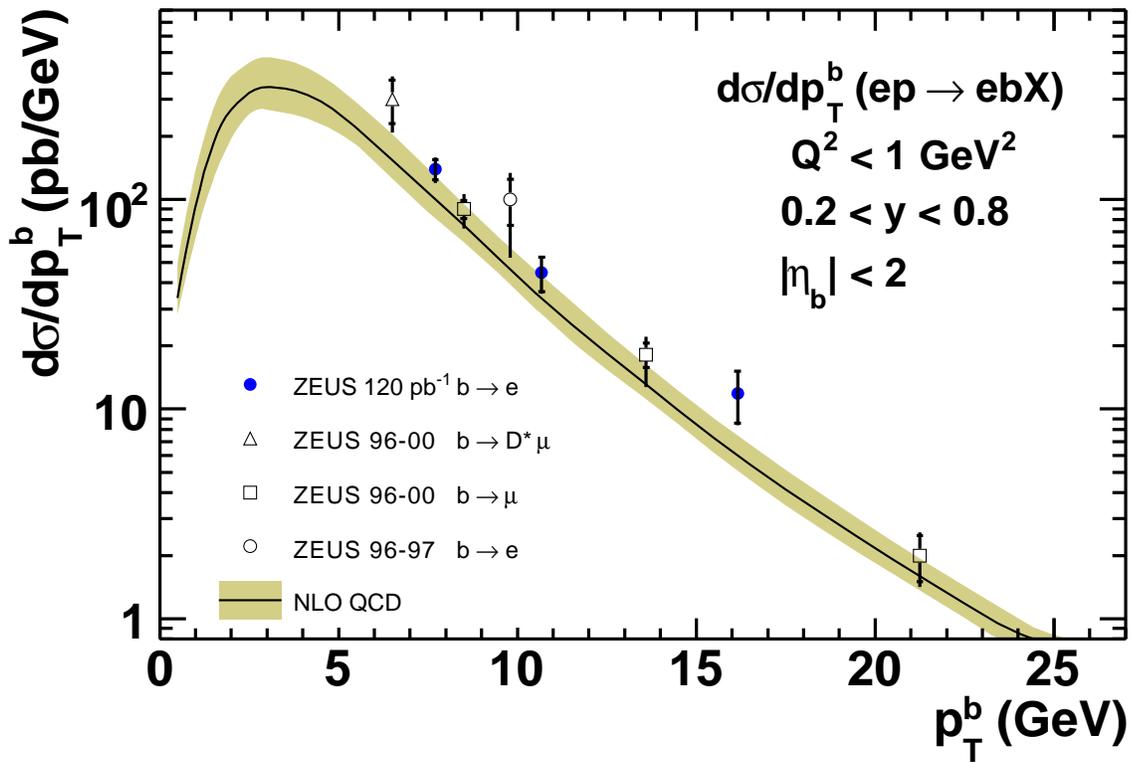}
  \caption{Differential cross section for $b$-quark production as a function of
    transverse momentum, \pTb, compared to the results of previous
    ZEUS measurements as indicated in the figure.
    The measurements are shown as points. The inner error bar shows the
    statistical uncertainty and the outer error bar shows the statistical and
    systematic uncertainties added in quadrature.
    The solid line shows the NLO QCD prediction from the FMNR program with the theoretical 
    uncertainty shown as the shaded band.}
  \label{fig:ptb}
  \vfill
\end{figure}

The differential cross sections as a function of the transverse energy
of the jet associated with the electron from the semileptonic decay,
\ETejet, were also determined. These cross sections are shown in
Fig.~\ref{fig:etb} and given in Table~\ref{tab:jetb}.
The good agreement with the NLO QCD prediction allows 
the cross section as a function of \pTb{} to be
extracted\cite{epj:c18:625}. 
The resulting cross section is shown in
Fig.~\ref{fig:ptb} and is also compared with previous ZEUS
measurements\cite{epj:c18:625,pr:d70:012008,epj:c50:1434}.
The results presented here overlap in \pTb{} with
these previous measurements and have comparable or smaller
uncertainties, giving a consistent picture of $b$-quark production in
$ep$ collisions in the photoproduction regime.

% ----------------------------------------------------------------------------
%       Conclusions
% ----------------------------------------------------------------------------
\section{Conclusions}
\label{sec:conclusions}

Beauty and charm production have been measured in dijet
photoproduction using semileptonic decays into electrons. 
The results were obtained by simultaneously extracting the $b$- and
$c$-quark cross sections using a likelihood ratio optimised for
$b$-quark production. One of the most important variables in the
likelihood was the \dEdx{} measurement in the central tracking detector.

The results were compared to both NLO QCD calculations as well as 
predictions from Monte Carlo models. The NLO QCD predictions are
consistent with the data. The Monte Carlo models describe well the 
shape of the differential distributions in the 
data. The results on
$b$-quark production are also in agreement with a previous less
precise ZEUS measurement using semileptonic decays into electrons.
Within the momentum range covered by previous ZEUS measurements using
decays into muons, good agreement is found.

The cross section as a function of the transverse momentum of the $b$
quarks has been measured over a wide range.  The measurements agree
well with the previous values, giving a consistent picture of
$b$-quark production in $ep$ collisions in the photoproduction regime,
and are well reproduced by the NLO QCD calculations.

\section*{Acknowledgements}
\label{sec:acknowledge}

We thank the DESY Directorate for their strong support and
encouragement. The remarkable achievements of the HERA machine group were
essential for the successful completion of this work.  The design,
construction and installation of the ZEUS detector have been made possible by
the effort of many people who are not listed as authors.

\vfill\eject

%%% Local Variables: 
%%% mode: latex
%%% TeX-master: "BtoeHeraI"
%%% End: 

%------------------------------------------------------------------------------
%       Bibliography
%------------------------------------------------------------------------------
\clearpage
{\raggedright
\providecommand{\etal}{et al.\xspace}
\providecommand{\coll}{Collab.\xspace}
\catcode`\@=11
\def\@bibitem#1{%
\ifmc@bstsupport
  \mc@iftail{#1}%
    {;\newline\ignorespaces}%
    {\ifmc@first\else.\fi\orig@bibitem{#1}}
  \mc@firstfalse
\else
  \mc@iftail{#1}%
    {\ignorespaces}%
    {\orig@bibitem{#1}}%
\fi}%
\catcode`\@=12

}
%------------------------------------------------------------------------------
%       Tables
%------------------------------------------------------------------------------
%-------------------------------------------------------------------------------
%       Tables
%-------------------------------------------------------------------------------
%
% Phantom spacing for tables
\newcommand{\phd}{\phantom{.}}
\newcommand{\pho}{\phantom{0}}

\begin{table}[p]
  \centering
  \begin{tabular}{c | r@{$\pm$}>{$}l<{$} r@{$^{+}_{-}$}>{$}l<{$} r |  r@{$\pm$}>{$}l<{$} r@{$^{+}_{-}$}>{$}l<{$} r }
    &\multicolumn{2}{c}{$\sigvisb$}
    &\multicolumn{2}{c}{$\signlob$}
    &\Chadb 
    &\multicolumn{2}{c}{$\sigvisc$}
    &\multicolumn{2}{c}{$\signloc$}
    &\Chadc \\
    &\multicolumn{2}{c}{(pb)}
    &\multicolumn{2}{c}{(pb)}
    &
    &\multicolumn{2}{c}{(pb)}
    &\multicolumn{2}{c}{(pb)}
    & \\  \hline
    96---97 & 101 & 18^{ +8}_{-9}   &  \pho 81 & ^{20}_{12}  &  0.81 
    & 253 & 58^{+44}_{-22}   &  360 & ^{160}_{100}  &  1.00 \\
    98---00 & 139 & 16^{ +11}_{-12} & \pho 88 & ^{22}_{13}  &  0.81 
    & 260 & 40^{ +45}_{-23} & 380 & ^{170}_{110} &  1.01 \\
    \hline
    96---00 & 125 & 11^{+10}_{-11}   & \pho 88 & ^{22}_{13} & 0.81 
    & 278 & 33^{+48}_{-24}  & 380 & ^{170}_{110} & 1.01 
  \end{tabular}
  \caption{Total cross sections for electrons 
    from $b$ or $c$ quarks in photoproduction events, 
    $\Qsq < 1\gev^{2}$ and $0.2 < y < 0.8$, with 
    at least two jets with $\ETjet > 7(6)\gev$, $|\etajet| < 2.5$ 
    and the subsequent semileptonic decay to an electron with $\pTe >
    0.9\gev$ and $|\etae| < 1.5$. The values are given 
    separately for $\sqrt{s}=300\gev$ (96---97) and $\sqrt{s}=318\gev$
    (98---00) as well as for the complete data 
    set (96---00) extrapolated to $\sqrt{s}=318\gev$.
    The first error is statistical and the second is systematic. In
    addition, the NLO QCD prediction and its uncertainty is given,
    after applying the appropriate hadronisation correction (\Chadb, \Chadc).}
  \label{tab:visxsection}
\end{table}

\begin{table}[p]
  \centering
  \begin{tabular}{r@{~:~}l  | r@{$\pm$}>{$}l<{$}
      r@{$^{+}_{-}$}>{$}l<{$} p{0.75cm} |
      r@{$\pm$}>{$}l<{$} r@{$^{+}_{-}$}>{$}l<{$} p{0.6cm} }
    \multicolumn{2}{c|}{\centering $p_{T}^{e}$} &
    \multicolumn{2}{c}{\centering $d\sigma_b/dp_{T}^{e}$} &
    \multicolumn{2}{c}{\centering $d\signlob/dp_{T}^{e}$} &\Chadb & 
    \multicolumn{2}{c}{\centering $d\sigma_c/dp_{T}^{e}$} &
    \multicolumn{2}{c}{\centering $d\signloc/dp_{T}^{e}$} &\Chadc\\ [-0.2em]
    
    \multicolumn{2}{c|}{(\gev)}&
    \multicolumn{2}{c}{(pb/\gev)}&
    \multicolumn{2}{c}{(pb/\gev)} &  & 
    \multicolumn{2}{c}{(pb/\gev)}&
    \multicolumn{2}{c}{(pb/\gev)} & \\

    \hline
    0.9&2.1 &  56.3    & 9.6^{+4.3}_{-5.0}        & 34   & ^{11}_{7}     &  0.78
    & 117      &  26^{+20}_{-10}          & 177  & ^{71}_{38}    &  1.02 \\
    2.1&3.3 &  24.0    & 3.7^{+1.8}_{-2.1}        & 16.8 & ^{5.9}_{3.5}  &  0.79 
    &  54.4    & 9.0^{+9.5}_{-4.8}        & 80   & ^{42}_{23}    &  0.98 \\
    3.3&4.5 &  11.9    & 2.6^{+0.9}_{-1.1}        & 9.9  & ^{3.6}_{2.3}  &  0.84   
    &  26.0    & 5.8^{+4.5}_{-2.3}        & 36   & ^{27}_{14}    &  0.99 \\
4.5&8.0 &  4.7     & 1.9^{+0.4}_{-0.4}        & 3.3  & ^{1.4}_{0.9}  &  0.94 
&  1.5     & 2.7^{+0.3}_{-0.1}        & 7.5  & ^{9.5}_{4.0}  &  0.99 \\
\multicolumn{12}{c}{~} \\ [-0.5em]

\multicolumn{2}{c|}{~$\eta^{e}$}&
\multicolumn{2}{c}{$d\sigma_b/d\eta^{e}$}&
\multicolumn{2}{c}{$d\signlob/d\eta^{e}$} &\Chadb &
\multicolumn{2}{c}{$d\sigma_c/d\eta^{e}$}&
\multicolumn{2}{c}{$d\signloc/d\eta^{e}$} &\Chadc  \\[-0.2em]

\multicolumn{2}{c|}{}&
\multicolumn{2}{c}{(pb)}&
\multicolumn{2}{c}{(pb)} & &
\multicolumn{2}{c}{(pb)}&
\multicolumn{2}{c}{(pb)} &  \\

\hline
-1.5&-0.5&  26.4 & 4.6^{+2.0}_{-2.4}   & 16.7 & ^{6.6}_{3.6} &  0.75 
         &  51   & 12^{+9}_{-4}        & 111  & ^{66}_{33}   &  0.98 \\
-0.5& 0.0& 53.4  & 9.1^{+4.1}_{-4.8}   & 39.5 & ^{13.8}_{8.3}&  0.81 
         & 152   & 25^{+26}_{-13}      & 192  & ^{100}_{53}  &  1.01 \\
0.0&0.5  & 57.7  & 11.6^{+4.4}_{-5.1}  & 41.9 & ^{13.9}_{9.0}&  0.82 
         & 187   & 36^{+33}_{-16}      & 165  & ^{82}_{43}   &  1.02 \\
0.5&1.5  & 42.4  & 8.7^{+3.2}_{-3.8}   & 28.1 & ^{10.1}_{6.3}&  0.84 
         & 36    & 24^{+6}_{-3}        & 90   & ^{51}_{26}   &  1.02
\end{tabular}
\caption{Differential electron cross sections as a function of $p_{T}^{e}$
  and $\eta^{e}$ for the complete data set.
  For further details see the caption of Table~\protect\ref{tab:visxsection}.
}
\label{tab:electron}
\end{table}

\begin{table}[p]
\centering
  \begin{tabular}{r@{~:~}l  | r@{$\pm$}>{$}l<{$}
      r@{$^{+}_{-}$}>{$}l<{$} p{0.75cm} |
  r@{$\pm$}>{$}l<{$} r@{$^{+}_{-}$}>{$}l<{$} p{0.6cm} }
\multicolumn{2}{c|}{\centering $\xgamobs$}&
\multicolumn{2}{c}{\centering $d\sigma_{b}/d\xgamobs$}&
\multicolumn{2}{c}{\centering $d\signlob/d\xgamobs$} &\Chadb & 
\multicolumn{2}{c}{\centering $d\sigma_{c}/d\xgamobs$} &
\multicolumn{2}{c}{\centering $d\signloc/d\xgamobs$} &\Chadc\\ [-0.2em]

\multicolumn{2}{c|}{}&
\multicolumn{2}{c}{(pb)}&
\multicolumn{2}{c}{(pb)} &  & 
\multicolumn{2}{c}{(pb)}&
\multicolumn{2}{c}{(pb)} & \\ 

\hline
0.00 &0.45     &  51 &  17^{+4}_{-5}          &    28 & ^{18}_{10}   &  1.07 
               &  70 &  35^{+12}_{-6}         &   122 & ^{108}_{56}  &  1.16 \\
0.45 &0.75     &  166  &  25^{ +13}_{-15}     &    81 & ^{50}_{28}   &  2.27 
               &  227 &  49^{ +40}_{-20}      &   216 & ^{178}_{85}  &  1.32 \\
0.75 &1.00     &  216  &  31^{ +17}_{-19}     &   166 & ^{47}_{30}   &  0.55 
               &  715 &  79^{ +124}_{-63}     &   920 & ^{370}_{190} &  0.90 \\

\end{tabular}
\caption{Differential cross sections as a function of
  \xgamobs{} for the complete data set.
  For further details see the caption of Table~\protect\ref{tab:visxsection}.}
  \label{tab:xg}
\end{table}

\begin{table}[p]
\centering
  \begin{tabular}{r@{~:~}l  | r@{$\pm$}>{$}l<{$}
      r@{$^{+}_{-}$}>{$}l<{$} p{0.75cm} |
  r@{$\pm$}>{$}l<{$} r@{$^{+}_{-}$}>{$}l<{$} p{0.6cm} }
\multicolumn{2}{c|}{\ETjeti}&
\multicolumn{2}{c}{$d\sigma_b/\ETjeti$} &
\multicolumn{2}{c}{$d\signlob/\ETjeti$} &\Chadb & 
\multicolumn{2}{c}{$d\sigma_c/\ETjeti$} &
\multicolumn{2}{c}{$d\signloc/\ETjeti$} &\Chadc\\ [-0.2em]

\multicolumn{2}{c|}{(\gev)}&
\multicolumn{2}{c}{(pb/\gev)}&
\multicolumn{2}{c}{(pb/\gev)} &  & 
\multicolumn{2}{c}{(pb/\gev)}&
\multicolumn{2}{c}{(pb/\gev)} & \\ 
\hline
 7&10    & 16.8 &  2.5^{+1.3}_{-1.5}   & 10.1 & ^{3.2}_{1.9}  &  0.59 
         & 45.9 &  7.3^{+8.0}_{-4.0}   & 72   & ^{43}_{19}    &  0.99 \\
10&13    & 12.0 &   1.9^{+0.9}_{-1.1}  & 9.4  & ^{3.7}_{2.3}  &  0.97 
         &  28.0 &  4.7^{+4.9}_{-2.4}  & 35   & ^{14}_{12}    &  1.07 \\
13&16    & 8.3  &   1.6^{+0.6}_{-0.7}  & 5.1  & ^{2.0}_{1.1}  &  1.18
         &  5.9 &   3.4^{+1.0}_{-0.5}  & 11.7 & ^{7.0}_{2.9}  &  1.03 \\
16&30    & 1.00 &  0.38^{+0.08}_{-0.09}& 1.00 & ^{0.39}_{0.08}&  1.22
         &   1.5 &   1.1^{ +0.3}_{-0.1}& 1.8  & ^{1.2}_{0.5}  &  0.89 \\
\multicolumn{12}{c}{~} \\ [-0.5em]

\multicolumn{2}{c|}{\etajeti}&
\multicolumn{2}{c}{$d\sigma_b/d\etajeti$}&
\multicolumn{2}{c}{$d\signlob/d\etajeti$} &\Chadb &
\multicolumn{2}{c}{$d\sigma_c/d\etajeti$}&
\multicolumn{2}{c}{$d\signloc/d\etajeti$} &\Chadc  \\[-0.2em]

\multicolumn{2}{c|}{}&
\multicolumn{2}{c}{(pb)}&
\multicolumn{2}{c}{(pb)} & &
\multicolumn{2}{c}{(pb)}&
\multicolumn{2}{c}{(pb)} &  \\
\hline

 -1.0&-0.25 & 24.9 &  5.2^{ +1.9}_{-2.2}   & 17.5 & ^{6.1}_{2.7}&  0.82
            & 73   &  14^{+13}_{-6}        & 99   & ^{64}_{26}  &  0.95\\
-0.25& 0.5  & 47.6 &  8.2^{+3.7}_{-4.2}    & 42.6 & ^{12.7}_{7.7}&  1.01
            &177   &  24^{+31}_{-15}       & 164 & ^{75}_{35}   &  1.05\\
 0.5 & 1.5  & 49.3 &  7.8^{+3.8}_{-4.4}    & 30.4 & ^{7.9}_{6.1}&  0.91
            &71    &  17^{+12}_{-6}        & 106  & ^{41}_{32}  &  1.04\\
 1.5 & 2.5  & 23.7 &  5.5^{+1.8}_{-2.1}    &  9.2 & ^{3.6}_{2.4}&  0.76
            & 8    &  15^{ +1}_{-1}        & 35   & ^{23}_{12}  &  1.01
\end{tabular}
\caption{Differential cross sections for the most energetic jet
  as a function of \ETjet{} and \etajet{} for the complete data set.
  For further details see the caption of Table~\protect\ref{tab:visxsection}.}
  \label{tab:jet1}
\end{table}

\begin{table}[p]
\centering
  \begin{tabular}{r@{~:~}l  | r@{$\pm$}>{$}l<{$}
      r@{$^{+}_{-}$}>{$}l<{$} p{0.75cm} |
  r@{$\pm$}>{$}l<{$} r@{$^{+}_{-}$}>{$}l<{$} p{0.6cm} }
\multicolumn{2}{c|}{\centering $\ETejet $}&
\multicolumn{2}{c}{\centering $d\sigma_b/\ETejet$} &
\multicolumn{2}{c}{\centering $d\signlob/\ETejet$} &\Chadb & 
\multicolumn{2}{c}{\centering $d\sigma_c/\ETejet$} &
\multicolumn{2}{c}{\centering $d\signloc/\ETejet$} &\Chadc\\ [-0.2em]

\multicolumn{2}{c|}{(\gev)}&
\multicolumn{2}{c}{(pb/\gev)}&
\multicolumn{2}{c}{(pb/\gev)} &  & 
\multicolumn{2}{c}{(pb/\gev)}&
\multicolumn{2}{c}{(pb/\gev)} & \\ 
\hline
 6&10    & 16.1 &   1.8^{+1.2}_{-1.4}  &  12.3  & ^{5.1}_{3.0}   &  0.67
         & 42.2 &   5.2^{+7.3}_{-3.7}  &  64    & ^{38}_{18}     &  1.00\\
10&15    & 6.6  &   1.3^{+0.5}_{-0.6}  &  5.4   & ^{1.8}_{1.1}   &  1.00
         & 22.3 &   4.2^{+3.9}_{-2.0}  &  19.6  & ^{7.5}_{5.5}   &  1.06 \\
15&30    & 2.1  &   0.6^{+0.2}_{-0.2}  &   1.08 & ^{0.40}_{0.26} &  1.21
         & 0.3  &   1.9^{+0.1}_{-0.1}  &  1.7   & ^{1.2}_{0.5}   &  0.87
\end{tabular}
\caption{Differential cross sections of \ETejet{} for 
  the jet associated to the electron from beauty or charm decays for the
  complete data set.
  For further details see the caption of Table~\protect\ref{tab:visxsection}.}
  \label{tab:jetb}
\end{table}

%%% Local Variables: 
%%% mode: latex
%%% TeX-master: "BtoeHeraI"
%%% End: 

%------------------------------------------------------------------------------
%       Figures
%------------------------------------------------------------------------------
%------------------------------------------------------------------------------
%       Figures
%------------------------------------------------------------------------------

\clearpage 

\clearpage 

\clearpage 

\clearpage 

\clearpage 

\clearpage 

\clearpage 

\clearpage 

\clearpage 

\clearpage

\clearpage

\clearpage

\clearpage 

%%% Local Variables: 
%%% mode: latex
%%% TeX-master: "BtoeHeraI"
%%% End: 

%
%       ... that's it
%
\end{document}